\documentclass[12pt]{article}
\usepackage{axodraw}
\usepackage{graphicx}
\begin{document}
\baselineskip 0.6cm

\def\simgt{\mathrel{\lower2.5pt\vbox{\lineskip=0pt\baselineskip=0pt
           \hbox{$>$}\hbox{$\sim$}}}}
\def\simlt{\mathrel{\lower2.5pt\vbox{\lineskip=0pt\baselineskip=0pt
           \hbox{$<$}\hbox{$\sim$}}}}

\begin{titlepage}

\begin{flushright}
UCB-PTH-09/02
\end{flushright}

\vskip 1.0cm

\begin{center}

{\Large \bf 
Dark Matter Signals from Cascade Annihilations
}

\vskip 0.6cm

{\large
Jeremy Mardon, Yasunori Nomura, Daniel Stolarski, and Jesse Thaler
}

\vskip 0.4cm

{\it Berkeley Center for Theoretical Physics, Department of Physics, \\
  University of California, Berkeley, CA 94720 and} \\
{\it Theoretical Physics Group, Lawrence Berkeley National 
  Laboratory, Berkeley, CA 94720} \\

\vskip 0.8cm

\abstract{A leading interpretation of the electron/positron excesses 
 seen by PAMELA and ATIC is dark matter annihilation in the galactic 
 halo.  Depending on the annihilation channel, the electron/positron 
 signal could be accompanied by a galactic gamma ray or neutrino flux, 
 and the non-detection of such fluxes constrains the couplings and 
 halo properties of dark matter.  In this paper, we study the interplay 
 of electron data with gamma ray and neutrino constraints in the context 
 of cascade annihilation models, where dark matter annihilates into 
 light degrees of freedom which in turn decay into leptons in one 
 or more steps.  Electron and muon cascades give a reasonable fit to 
 the PAMELA and ATIC data.  Compared to direct annihilation, cascade 
 annihilations can soften gamma ray constraints from final state 
 radiation by an order of magnitude.  However, if dark matter annihilates 
 primarily into muons, the neutrino constraints are robust regardless 
 of the number of cascade decay steps.  We also examine the electron 
 data and gamma ray/neutrino constraints on the recently proposed 
 ``axion portal'' scenario.}

\end{center}
\end{titlepage}

\section{Introduction}
\label{sec:intro}

Recent observations by PAMELA~\cite{Adriani:2008zr} and 
ATIC~\cite{Chang:2008zz} strongly suggest a new primary source of 
galactic electrons and positrons.  Three leading interpretations of 
the PAMELA/ATIC excesses are astrophysical sources~\cite{Hooper:2008kg}, 
decay of dark matter~\cite{Yin:2008bs}, and annihilation of dark matter 
\cite{Cirelli:2008pk,ArkaniHamed:2008qn,Nomura:2008ru,Cholis:2008wq,%
Fairbairn:2008fb}.  While the current PAMELA/ATIC data cannot distinguish 
between these possibilities, one expects that the correct scenario will 
ultimately be determined with the help of complementary data 
from synchrotron, gamma ray, and neutrino telescopes, as well 
as collider and direct detection experiments.

One piece of data that points toward an annihilation interpretation 
is the WMAP Haze~\cite{Finkbeiner:2003im}, an apparent excess of 
synchrotron radiation coming from the galactic center.   Dark matter 
annihilation into charged particles is uniquely positioned to explain 
the Haze \cite{Finkbeiner:2004us,Cholis:2008vb}.  If $n$ is the dark 
matter number density near the galactic center, then the synchrotron 
signal for dark matter annihilation scales like $n^2$, while the signal 
for dark matter decay scales only as $n$.  (Astrophysical signals also 
roughly scale like $n$.)  Given the normalization of the PAMELA/ATIC 
excess, the $n^2$ scaling is favored to explain the size of the Haze 
anomaly~\cite{Zhang:2008tb}.

On the other hand, the same $n^2$ versus $n$ logic implies that the 
dark matter annihilation interpretation is more strongly constrained 
by the \emph{absence} of gamma ray or neutrino excesses from the 
galactic center.  While these constraints are dependent on the 
Milky Way dark matter halo profile, there are already strong 
bounds on the annihilation interpretation for strongly peaked halos 
\cite{Bell:2008vx,Bertone:2008xr,Bergstrom:2008ag,Hisano:2008ah,%
Liu:2008ci}.  Therefore, it is worth exploring dark matter annihilation 
scenarios in detail to understand how robust the tension is between 
explaining PAMELA/ATIC/Haze and satisfying other bounds.

Given the absence of anti-proton~\cite{Adriani:2008zq} or gamma 
ray \cite{Aharonian:2006wh,Aharonian:2006au,Aharonian:2007km} excesses, 
the dark matter annihilation scenarios favored to explain PAMELA/ATIC 
involve annihilation into electrons and muons.  However, dark matter 
need not annihilate into leptons directly.  There are a variety 
of ``cascade annihilation'' models where dark matter annihilates 
into light resonances which in turn decay into electrons or muons. 
These light resonances can lead to nonperturbative enhancements 
\cite{Hisano:2003ec,Pospelov:2008jd} of the dark matter annihilation 
rate in the galactic halo, providing the large boost factors necessary to 
explain PAMELA/ATIC \cite{Cirelli:2008pk,ArkaniHamed:2008qn,Nomura:2008ru}. 
Also, annihilation into light fields gives a kinematic explanation 
for why dark matter annihilation preferentially yields light 
leptons \cite{Cholis:2008vb,ArkaniHamed:2008qn,Nomura:2008ru}. 
Previous studies of cascade annihilation models appear in 
Refs.~\cite{Cholis:2008wq,Bergstrom:2008ag}.

In the present context, these cascade annihilation scenarios 
are interesting because they have the potential to explain 
PAMELA/ATIC while weakening constraints from gamma rays, 
as measured by atmospheric Cerenkov telescopes like H.E.S.S. 
\cite{Aharonian:2006wh,Aharonian:2006au,Aharonian:2007km}. 
The reason is that gamma ray experiments are directly sensitive 
to the primary injection spectra, and cascade annihilations yield 
softer and smaller injection spectra of gamma rays from final state 
radiation (FSR).  PAMELA/ATIC sees electrons and positrons through 
the filter of charged cosmic ray transport, a process which introduces 
large uncertainties.  Considering also the uncertainties in the 
highest energy ATIC data, we find that softer spectra of primary 
leptons can still explain the PAMELA/ATIC excesses.

For cascade annihilations that terminate in muons, there is also 
an irreducible source of galactic neutrinos, which can be observed 
as an upward-going muon flux on earth, for example, by water Cerenkov 
detectors like Super-Kamiokande (Super-K)~\cite{Desai:2004pq}.  While 
cascades soften the neutrino spectrum, we will see that the final 
constraints from neutrinos are rather insensitive to the number 
of cascade steps, and may provide the most robust bound on muon 
cascade scenarios.

The organization of this paper is as follows.  In the next section, 
we define our framework for analyzing signals of dark matter through 
cascade annihilations, with details of the cascade energy spectra given 
in Appendix~\ref{app:spectra}.  In Section~\ref{sec:PAM-ATIC}, we find 
the best fit dark matter masses and annihilation cross sections for 
various cascade scenarios given the PAMELA/ATIC data.  We consider 
H.E.S.S. gamma ray bounds from FSR in Section~\ref{sec:gamma} 
and Super-K neutrino bounds in Section~\ref{sec:neutrino}.  In 
Section~\ref{sec:axion}, we study a particular cascade annihilation 
scenario called the axion portal~\cite{Nomura:2008ru}, and present a 
less constrained ``leptonic'' version in Appendix~\ref{app:lepto-axion}. 
Conclusions are given in Section~\ref{sec:concl}.

\section{Cascade Annihilations}
\label{sec:cascade}

If dark matter is a thermal relic, then it will have at least one 
annihilation mode into standard model fields, since in the early universe, 
the dark matter annihilation channels keep dark matter in thermal 
equilibrium with the standard model until freezeout.  However, dark 
matter need not annihilate into standard model particles directly; 
it can annihilate into new (unstable) resonances which in turn decay 
into standard model fields.  As long as the new resonances are 
sufficiently broad, then dark matter will be in close enough thermal 
contact with the standard model for a freezeout calculation to 
be valid.

\begin{figure}[t]
\begin{center}
\begin{picture}(440,130)(50,-15)
%
%
  \Text(90,-10)[t]{\large direct}
  \Line(30,70)(90,50) \Text(29,70)[r]{$\chi$}
  \Line(30,30)(90,50) \Text(29,30)[r]{$\bar{\chi}$}
  \Line(90,50)(150,70) \Text(155,72)[l]{$\ell^+$}
  \Line(90,50)(150,30) \Text(155,30)[l]{$\ell^-$}
  \GCirc(90,50){10}{0.7}
%
%
  \Text(250,-10)[t]{\large $1$ step}
  \Line(190,65)(230,50) \Text(189,65)[r]{$\chi$}
  \Line(190,35)(230,50) \Text(189,35)[r]{$\bar{\chi}$}
  \Line(230,50)(270,70) \Text(261,67)[br]{$\phi_1$}
  \Line(230,50)(270,30) \Text(261,36)[tr]{$\phi_1$}
  \GCirc(230,50){10}{0.7}
  \Line(270,70)(310,80) \Text(315,82)[l]{$\ell^+$}
  \Line(270,70)(310,60) \Text(315,60)[l]{$\ell^-$}
  \Line(270,30)(310,40) \Text(315,42)[l]{$\ell^+$}
  \Line(270,30)(310,20) \Text(315,20)[l]{$\ell^-$}
%
%
  \Text(410,-10)[t]{\large $2$ step}
  \Line(350,65)(380,50) \Text(349,65)[r]{$\chi$}
  \Line(350,35)(380,50) \Text(349,35)[r]{$\bar{\chi}$}
  \Line(380,50)(410,70) \Text(405,67)[br]{$\phi_2$}
  \Line(380,50)(410,30) \Text(405,36)[tr]{$\phi_2$}
  \GCirc(380,50){10}{0.7}
  \Line(410,70)(440,85) \Text(435,83)[br]{$\phi_1$}
  \Line(410,70)(440,60) \Text(435,63)[tr]{$\phi_1$}
  \Line(410,30)(440,40) \Text(435,39)[br]{$\phi_1$}
  \Line(410,30)(440,15) \Text(435,20)[tr]{$\phi_1$}
  \Line(440,85)(470,90) \Text(475,93)[l]{$\ell^+$}
  \Line(440,85)(470,80) \Text(475,81)[l]{$\ell^-$}
  \Line(440,60)(470,65) \Text(475,68)[l]{$\ell^+$}
  \Line(440,60)(470,55) \Text(475,56)[l]{$\ell^-$}
  \Line(440,40)(470,45) \Text(475,46)[l]{$\ell^+$}
  \Line(440,40)(470,35) \Text(475,34)[l]{$\ell^-$}
  \Line(440,15)(470,20) \Text(475,21)[l]{$\ell^+$}
  \Line(440,15)(470,10) \Text(475,9)[l]{$\ell^-$}
  \Text(500,50)[l]{\large $\cdots$}
\end{picture}
\end{center}
\caption{For an $n$-step cascade annihilation, dark matter $\chi$ 
 annihilates into $\phi_n \phi_n$.  The cascade annihilation then occurs 
 through $\phi_{i+1} \rightarrow \phi_i \phi_i$ ($i=1,\cdots,n-1$), 
 and in the last stage, $\phi_1$ decays into standard model 
 particles.  The figure represents the cases where $\phi_1 \rightarrow 
 \ell^+ \ell^-$.}
\label{fig:cascade}
\end{figure}
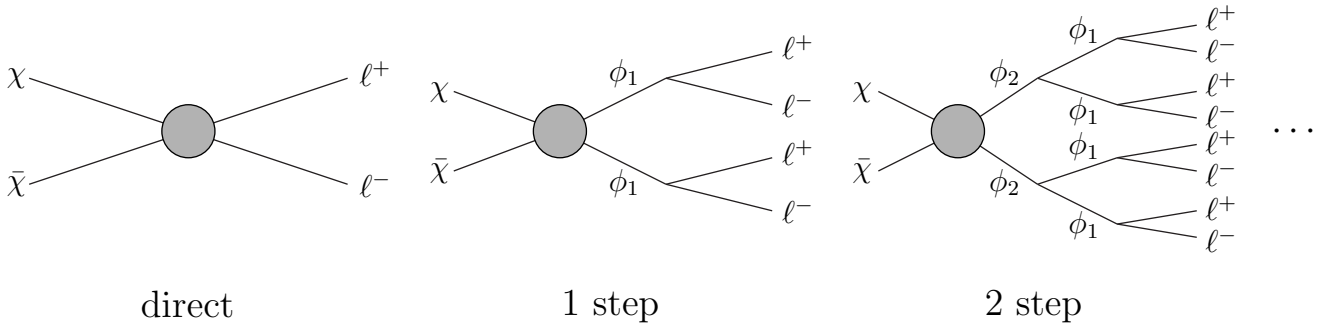
These ``cascade annihilations'' can occur in one or more steps, as shown 
in Figure~\ref{fig:cascade}, and there are a variety of motivations for 
such cascade scenarios.  Since the direct coupling between dark matter 
and the standard model can be small while still achieving the desired 
thermal relic abundance, such models can have reduced direct detection 
cross sections compared to generic weakly interacting massive particles 
(WIMPs)~\cite{Pospelov:2007mp}.  Cascade annihilations can arise 
whenever new light resonances have large couplings to dark matter, 
such as in exciting dark matter~\cite{Finkbeiner:2007kk}.  In the 
context of PAMELA/ATIC, there has been interest in using light 
resonances to provide large enhancement to the galactic annihilation 
rate through the Sommerfeld effect or bound state (WIMPonium) 
formation \cite{Cirelli:2008pk,ArkaniHamed:2008qn,Nomura:2008ru}. 
These light resonances can also explain the lepton-richness of 
dark matter annihilation through kinematic thresholds 
\cite{ArkaniHamed:2008qn,Nomura:2008ru}.

Cascade annihilations give softer primary spectra for the 
annihilation products than direct annihilation.  As reviewed in 
Appendix~\ref{app:spectra}, for (scalar) cascade annihilations 
involving well-separated kinematic scales, the approximate primary 
spectra can be calculated using a simple convolution formula.  The 
energy spectra are conveniently parametrized in terms of the energy 
fraction $x_n \equiv E_n/m_{\rm DM}$, where $E_n$ is the final state 
energy after the $n$-th step of the cascade, and $m_{\rm DM}$ is the 
dark matter mass.  If $d\tilde{N}/dx_0$ is the normalized primary 
spectrum from direct annihilation, then the normalized primary 
spectrum from a $1$-step annihilation $d\tilde{N}/dx_1$ is
\begin{equation}
  \frac{d \tilde{N}}{d x_1} \approx \int_{x_1}^1\! \frac{d x_0}{x_0}\, 
    \frac{d \tilde{N}}{d x_0}.
\label{eq:simp-conv}
\end{equation}
This convolution formula can be iterated to obtain the normalized primary 
spectrum for an $n$-step cascade scenario.  Eq.~(\ref{eq:simp-conv}) is 
also reasonably representative of non-scalar cascades, since indirect 
detection measurements are unpolarized.

While cascade annihilations give softer spectra, they typically yield 
a higher multiplicity of annihilation products.  The final primary 
spectrum $dN/dx_n$ must take into account the multiplicity of annihilation 
products per dark matter annihilation, and in a typical $n$-step cascade, 
the final state multiplicity scales like $2^n$.

An important exception, however, is in cases involving singularities, 
where cascade annihilations yield both a softer spectrum {\it and} 
a lower multiplicity.  For example, FSR from charged leptons has 
a collinear singularity that is regulated by the lepton mass. 
Integrating over the singularity gives a photon spectrum $dN_\gamma/dx$ 
that is proportional to $\ln Q/m_\ell$, where $Q$ is the total energy 
available for radiation.  For direct annihilation $Q \simeq 2 m_{\rm DM}$, 
while for cascade annihilations $Q \simeq m_1$, where $m_1$ is the mass 
of the resonance in the last stage of the annihilation $\phi_1 \rightarrow 
\ell^+ \ell^-$.  Light enough $\phi_1$ fields can give a dramatic 
reduction in the $\ln Q/m_\ell$ factor and thus the FSR photon yield.

In this paper, we only consider cascade annihilations that terminate 
in electrons or muons.  The same analysis, however, could be repeated 
for charged pions or taus by changing the direct annihilation spectra. 
We expect the results for an $n$-step charged pion cascade to be similar 
to an $(n+1)$-step muon cascade.  Cascades involving taus will face 
stronger gamma ray bounds because of the $O(1)$ fraction of $\pi^0$s 
in tau decays.  For simplicity, we only show plots for direct, $1$-step, 
and $2$-step cascades.

\section{PAMELA/ATIC Spectra}
\label{sec:PAM-ATIC}

The PAMELA satellite experiment~\cite{Picozza:2006nm} observed 
an anomalous source of galactic positrons in the energy range 
$10$--$100~{\rm GeV}$ through a measurement of the positron fraction 
$\Phi_{e^+}/(\Phi_{e^+} + \Phi_{e^-})$~\cite{Adriani:2008zr}.  The 
ATIC balloon experiment~\cite{Guzik:2004} is not capable of charge 
separation, but observed a peak in the total electron plus positron 
flux between $100$--$1000~{\rm GeV}$ in a $(\Phi_{e^+} + \Phi_{e^-})$ 
measurement~\cite{Chang:2008zz}.  Intriguingly, both excesses 
can be described by a single new source of galactic electrons and 
positrons, and here we study the goodness of fit for dark matter 
cascade annihilations.  The primary electron/positron spectra 
for $n$-step electron and muon cascade annihilations are given 
in Appendices~\ref{subapp:direct-e} and \ref{subapp:muon-decay}. 
Once the primary spectrum is known and a dark matter halo profile 
assumed, we can propagate the electrons and positrons through 
the Milky Way and compare with the PAMELA/ATIC data.

We follow the analysis of Ref.~\cite{Delahaye:2007fr}, which 
assumes that galactic electrons and positrons can be described by 
a diffusion-loss process.  In the turbulent galactic magnetic fields, 
electrons/positrons diffuse within a fiducial region around the galactic 
disk and escape the galaxy outside that region.  An energy loss term 
incorporates the physics of inverse Compton scattering (ICS) and 
synchrotron radiation.  Dark matter annihilation is represented by 
a source term proportional to the square of the dark matter halo 
density.  Since the energy loss time is much shorter than the age 
of the galaxy, the electron/positron system is assumed to be in 
steady state.

Taking $\psi_{e^-}(\vec{x},E)$ to be the galactic electron number 
density per unit energy, the diffusion-loss equation is
\begin{equation}
  K_0\, \varepsilon^\delta\, \nabla^2 \psi_{e^-}(\vec{x},E) 
    + \frac{\partial}{\partial\varepsilon} 
      \left( \frac{\varepsilon^2}{\tau_E}\, \psi_{e^-}(\vec{x},E) \right) 
    + q(\vec{x},E) = 0,
\label{eq:prop-eq}
\end{equation}
where $\varepsilon = E/{\rm GeV}$, $K_0$ and $\delta$ parametrize 
the (energy dependent) diffusion, $\tau_E$ is a characteristic 
energy loss time, and $q(\vec{x},E)$ is the electron source term 
for dark matter annihilations.  The same equation also holds for 
the positron number density per unit energy $\psi_{e^+}(\vec{x},E)$. 
The electron/positron densities $\psi_{e^-/e^+}(\vec{x},E)$ are 
assumed to have vanishing boundary conditions on the surface of 
a cylinder of height $2L$ and radius $R$.   We consider the three 
benchmark models from Ref.~\cite{Delahaye:2007fr}, which are 
summarized in Table~\ref{tab:diffloss}.
\begin{table}
\begin{center}
\begin{tabular}{c|ccccc}
  & $R$ (kpc) & $L$ (kpc) & $K_0$ (kpc$^2$/Myr) & $\delta$ & $\tau_E$ (sec)
\\ \hline
  MED & $20$ &  $4$ &  $0.0112$ & $0.70$ & $10^{16}$ \\
  M1  & $20$ & $15$ &  $0.0765$ & $0.46$ & $10^{16}$ \\
  M2  & $20$ &  $1$ & $0.00595$ & $0.55$ & $10^{16}$ 
\end{tabular}
\end{center}
\caption{Diffusion-loss parameters for the three benchmark models 
 (MED, M1, and M2) for electron/positron propagation.}
\label{tab:diffloss}
\end{table}

The electron/positron source term is given by
\begin{equation}
  q(\vec{x},E) = \frac{1}{2\eta} \langle \sigma v \rangle 
    \left(\frac{\rho(\vec{x})}{m_{\rm DM}} \right)^2 \frac{d N_e}{d E},
\label{eq:source-term}
\end{equation}
where $\rho(\vec{x})$ is an assumed dark matter halo profile, $m_{\rm DM}$ 
is the dark matter mass, $\langle \sigma v \rangle$ is the average dark 
matter annihilation cross section in the galactic halo, and $dN_e/dE$ 
is the electron energy spectrum per dark matter annihilation.  $\eta = 1$ 
if dark matter is self-conjugate (e.g.\ a Majorana fermion), while 
$\eta = 2$ if not (e.g.\ a Dirac fermion).  We consider three spherically 
symmetric benchmark halo profiles (cored isothermal~\cite{Bahcall:1980fb}, 
NFW~\cite{Navarro:1996gj}, and Einasto~\cite{Navarro:2003ew}) with 
$r_\odot = 8.5~{\rm kpc}$ and $\rho_\odot = 0.3~{\rm GeV}\, {\rm cm}^{-3}$:
\begin{eqnarray}
  && \rho(r)_{\rm Isothermal} 
  = \rho_\odot \frac{1 + (r_\odot/r_c)^2}{1 + (r/r_c)^2}, 
  \qquad r_c = 5~{\rm kpc},
\label{eq:rho-Isothermal}\\
  && \rho(r)_{\rm NFW} 
  = \rho_\odot \frac{r_\odot}{r} 
    \left(\frac{1 + r_\odot/r_c}{1 + r/r_c} \right)^2, 
  \qquad r_c = 20~{\rm kpc},
\label{eq:rho-NFW}\\
  && \rho(r)_{\rm Einasto} 
  = \rho_\odot \exp\left\{ -\frac{2}{\alpha} 
    \left[ \left(\frac{r}{r_c}\right)^\alpha 
    - \left(\frac{r_\odot}{r_c}\right)^\alpha \right] \right\}, 
  \qquad \alpha = 0.17, \quad r_c = 20~{\rm kpc}.
\label{eq:rho-Einasto}
\end{eqnarray}
$N$-body simulations suggest that Einasto and NFW are more realistic 
profiles for $r \simgt 1~{\rm kpc}$.  Within the inner region, however, 
there is considerable uncertainty, and we include the cored isothermal 
profile to explore the possibility of a less peaked distribution.

Once the source term is specified, Eq.~(\ref{eq:prop-eq}) can be 
solved using the methods of \cite{Delahaye:2007fr,Hisano:2005ec}, 
and the electron/positron intensities (fluxes per energy per solid 
angle) at the earth due to dark matter annihilations are given by
\begin{equation}
  \frac{d\Phi_{e^-/e^+}^{(\rm DM)}}{dE\, d\Omega}(E) 
  = \frac{B_{e,{\rm astro}}}{4\pi} \psi_{e^-/e^+}(\vec{x}_\odot,E),
\label{eq:e-flux-DM}
\end{equation}
where $\vec{x}_\odot$ is the location of the solar system and 
$B_{e,{\rm astro}}$ is an astrophysical boost factor.  Since the 
annihilation rate is proportional to the squared density of dark matter 
$\rho_{\rm DM}^2$, clumping of dark matter tends to increase the local 
electron/positron flux, and the $B_{e,{\rm astro}}$ factor accounts for 
differences from the assumed smooth halo profile, as well as uncertainties 
in $\rho_\odot$.  Of course, in the presence of clumpiness, the true 
electron/positron spectrum is also modified~\cite{Lavalle:2006vb}.

Galactic cosmic rays are a known source of background primary electrons. 
Background secondary electrons and positrons arise, e.g., from 
collisions of cosmic ray protons with interstellar gas.  In principle, 
the spectra of background electrons/positrons are correlated with the 
diffusion-loss parameters for electrons/positrons, but for simplicity 
we will use the parameterization of background primaries and 
secondaries from Ref.~\cite{Moskalenko:1997gh}:
\begin{eqnarray}
  \frac{d\Phi_{e^-}^{(\rm prim)}}{dE\, d\Omega} 
  &=& \frac{0.16\,\varepsilon^{-1.1}}{1 + 11\,\varepsilon^{0.9} 
    + 3.2\,\varepsilon^{2.15}}\,\,\, 
    {\rm GeV}^{-1}\, {\rm cm}^{-2}\, {\rm s}^{-1}\, {\rm sr}^{-1},
\label{eq:Phi_e-_prim}\\
  \frac{d\Phi_{e^-}^{(\rm sec)}}{dE\, d\Omega} 
  &=& \frac{0.70\,\varepsilon^{0.7}}{1 + 110\,\varepsilon^{1.5} 
    + 600\,\varepsilon^{2.9} + 580\,\varepsilon^{4.2}}\,\,\, 
    {\rm GeV}^{-1}\, {\rm cm}^{-2}\, {\rm s}^{-1}\, {\rm sr}^{-1},
\label{eq:Phi_e-_sec}\\
  \frac{\Phi_{e^+}^{(\rm sec)}}{dE\, d\Omega} 
  &=& \frac{4.5\,\varepsilon^{0.7}}{1 + 650\,\varepsilon^{2.3} 
    + 1500\,\varepsilon^{4.2}}\,\,\,
    {\rm GeV}^{-1}\, {\rm cm}^{-2}\, {\rm s}^{-1}\, {\rm sr}^{-1},
\label{eq:Phi_e+_sec}
\end{eqnarray}
where again $\varepsilon = E/{\rm GeV}$.  To treat background 
uncertainties, we will marginalize over the normalization and 
overall slope of the background in our analysis:
\begin{equation}
  \Phi_{e^-}^{(\rm back.~fit)} = A_- \varepsilon^{P_-} 
    \left( \Phi_{e^-}^{(\rm prim)} + \Phi_{e^-}^{(\rm sec)} \right),
\qquad
  \Phi_{e^+}^{(\rm back.~fit)} = A_+ \varepsilon^{P_+} 
    \left( \Phi_{e^+}^{(\rm sec)} \right),
\label{eq:marginalize}
\end{equation}
where we allow $0 < A_\pm < \infty$, $-0.05 < P_\pm < 0.05$, as 
in~\cite{Cirelli:2008pk}.

In the limit that Eq.~(\ref{eq:simp-conv}) holds, any given dark 
matter cascade topology has just two free parameters: the dark 
matter mass $m_{\rm DM}$ and the annihilation cross section $\langle 
\sigma v \rangle$.  Following the literature, we normalize the cross 
section to the value that leads to the right relic thermal abundance 
$\langle \sigma v \rangle_0 \simeq 3 \times 10^{-26}\, \eta\, {\rm cm}^3\, 
{\rm s}^{-1}$ and express our results in terms of an effective boost 
factor
\begin{equation}
  B = B_{e,{\rm astro}} 
    \frac{\langle \sigma v \rangle}{\langle \sigma v \rangle_0},
\label{eq:def-B}
\end{equation}
which includes both the deviation from the naive thermal freezeout cross 
section and dark matter clumping.  Using only the statistical error bars, 
we perform a chi-squared fit of the derived electron/positron intensities 
to the PAMELA $e^+/(e^+ + e^-)$ and ATIC $e^+ + e^-$ data, treating 
$m_{\rm DM}$ and $B$ as free parameters.  We use only $E \simgt 
10~{\rm GeV}$ bins for the PAMELA data, as the lower energy bins 
are strongly affected by solar modulation effects (see the discussion 
in~\cite{Adriani:2008zr}).

\begin{figure}
  \center{\includegraphics[scale=0.98]{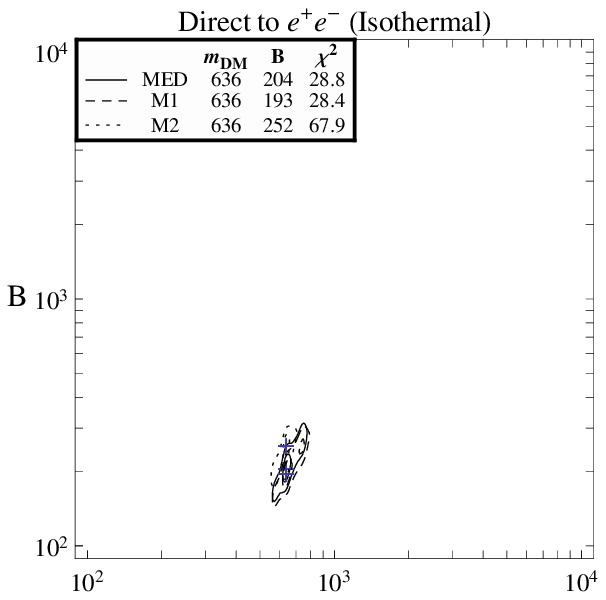}
  \includegraphics[scale=0.98]{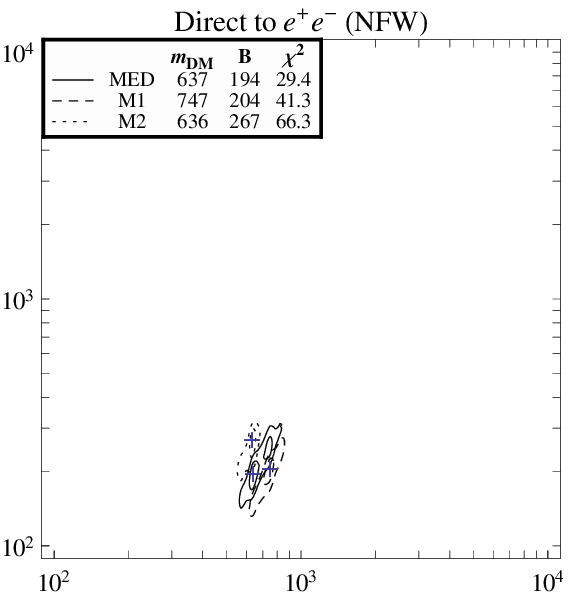}
  \includegraphics[scale=0.98]{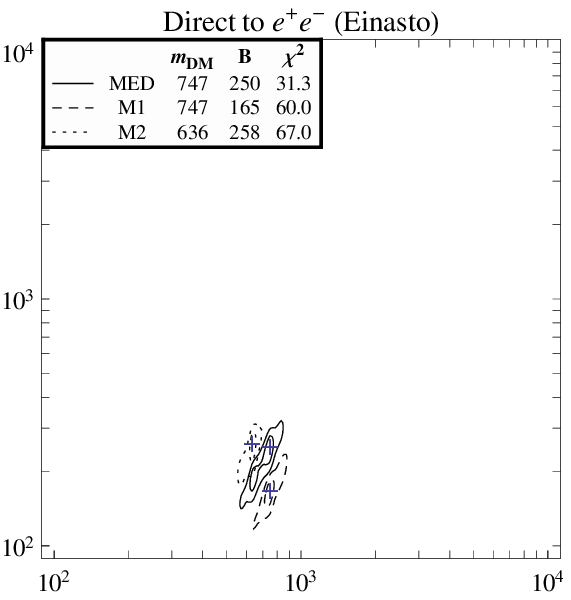}}
  \center{\includegraphics[scale=0.98]{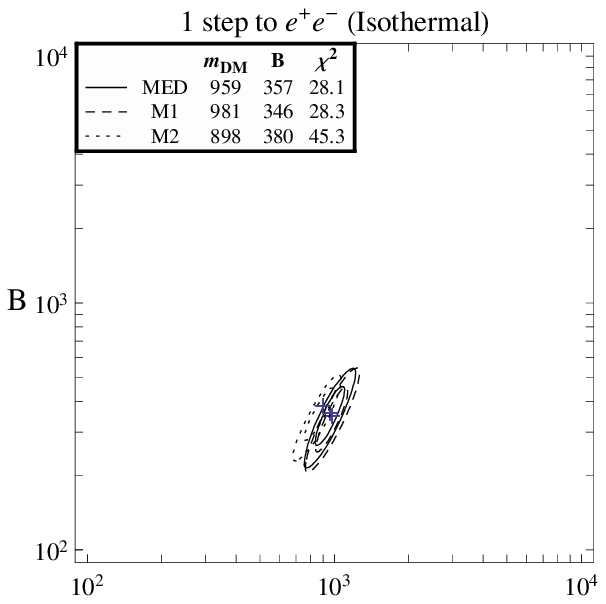}
  \includegraphics[scale=0.98]{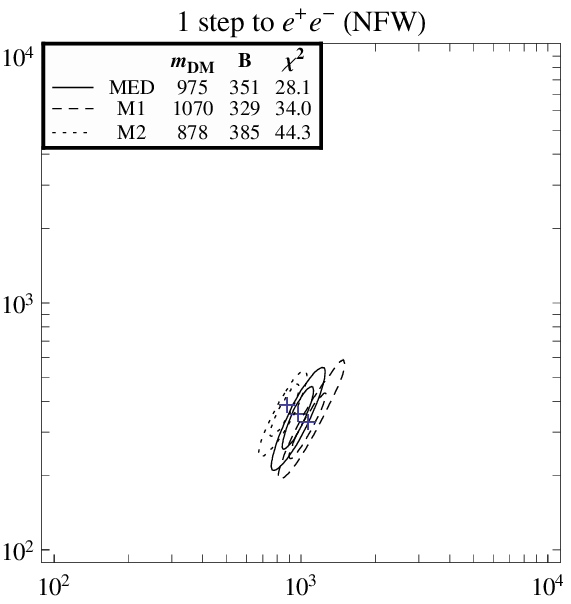}
  \includegraphics[scale=0.98]{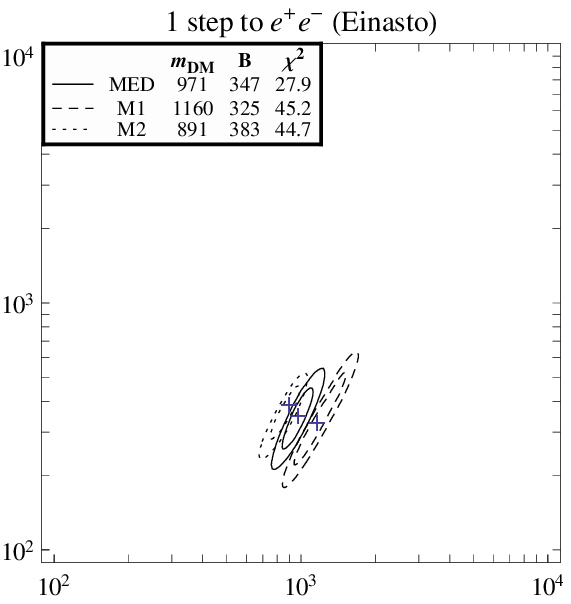}}
  \center{\includegraphics[scale=0.98]{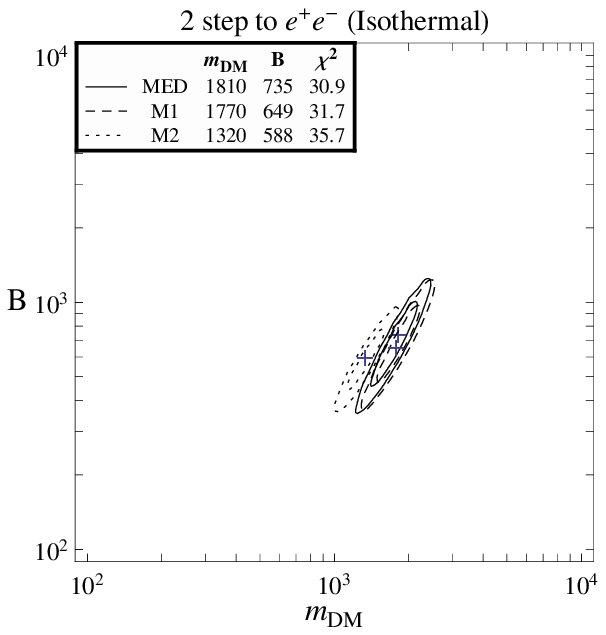}
  \includegraphics[scale=0.98]{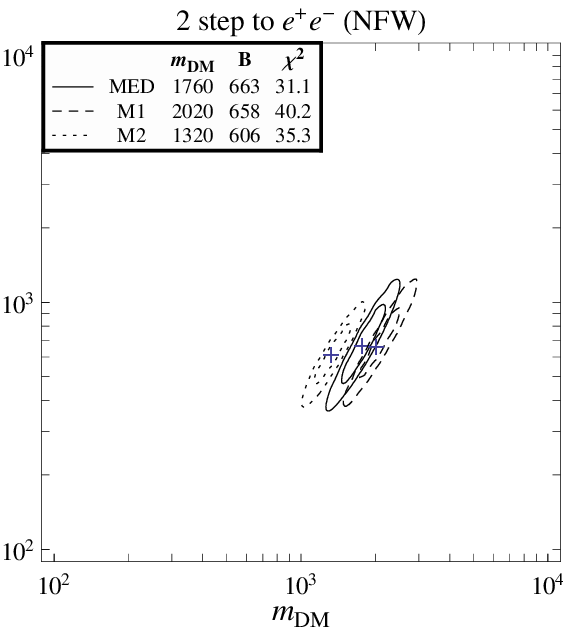}
  \includegraphics[scale=0.98]{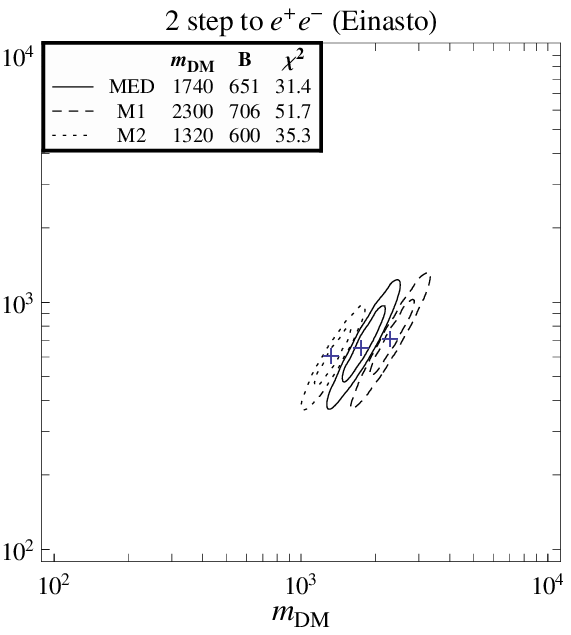}}
\caption{The best fit regions for the dark matter mass $m_{\rm DM}$ 
 and boost factor $B$ in the cases of direct, $1$-step, and $2$-step 
 annihilations into $e^+ e^-$ for different halo profiles and propagation 
 models.  The best fit values are indicated by the crosses, and the 
 contours are for $1\sigma$ and $2\sigma$.}
\label{fig:positron_e}
\end{figure}
\begin{figure}
  \center{\includegraphics[scale=0.98]{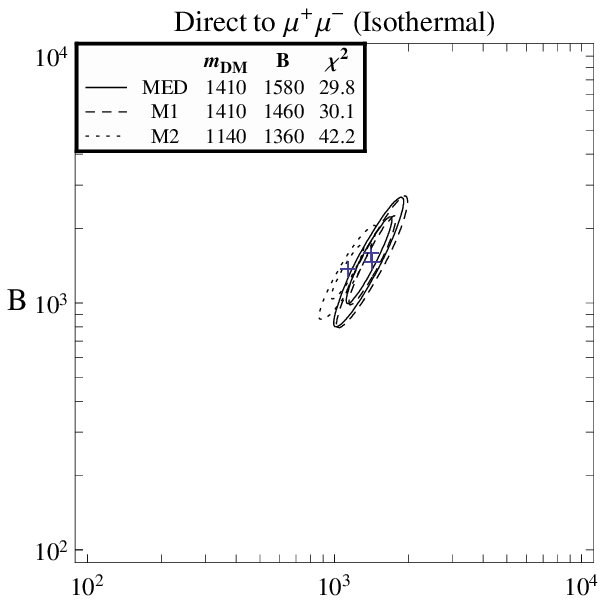}
  \includegraphics[scale=0.98]{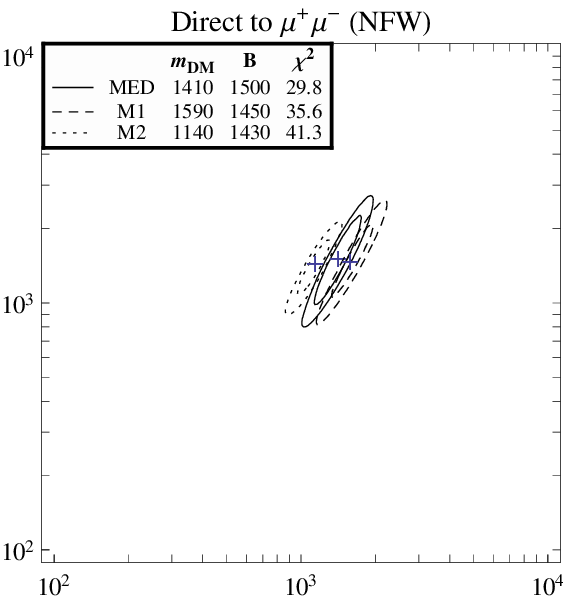}
  \includegraphics[scale=0.98]{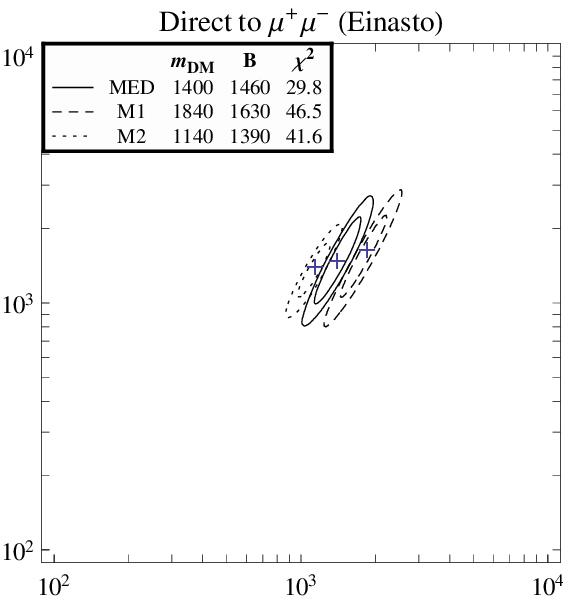}}
  \center{\includegraphics[scale=0.98]{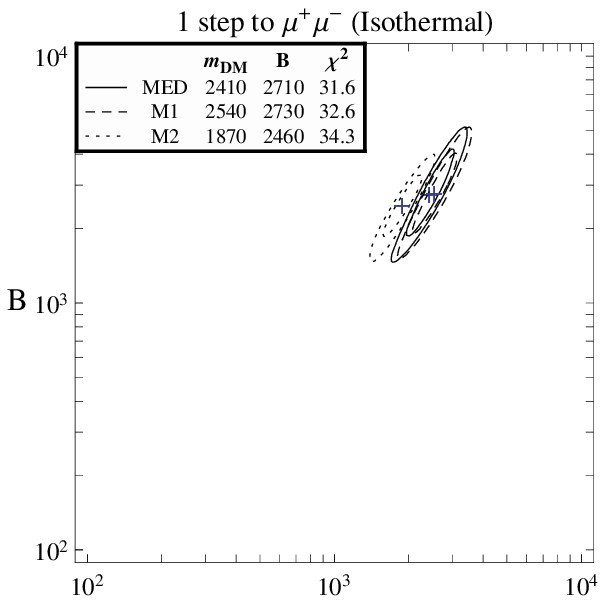}
  \includegraphics[scale=0.98]{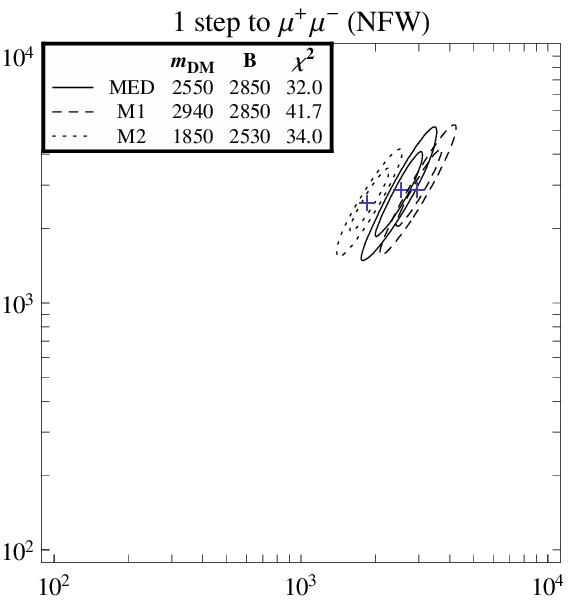}
  \includegraphics[scale=0.98]{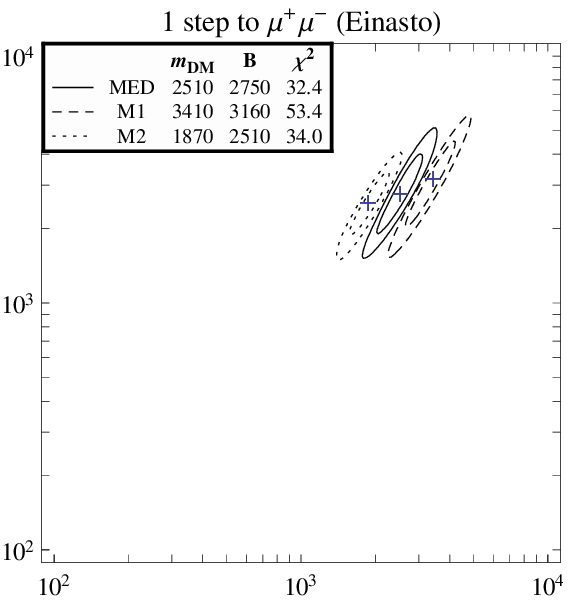}}
  \center{\includegraphics[scale=0.98]{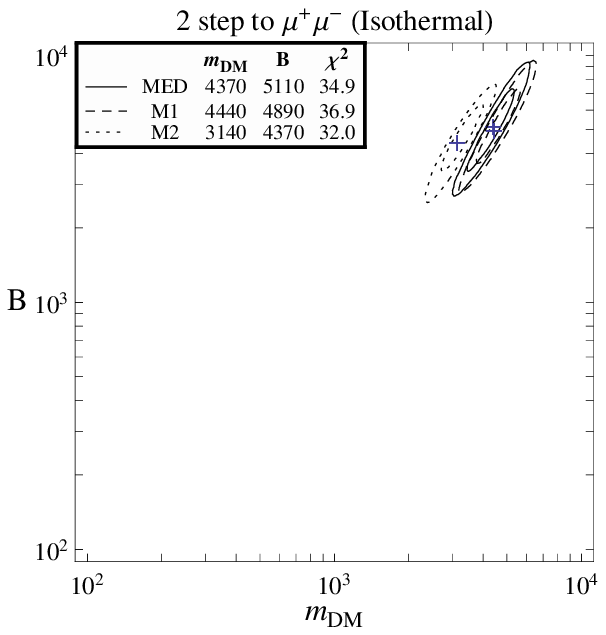}
  \includegraphics[scale=0.98]{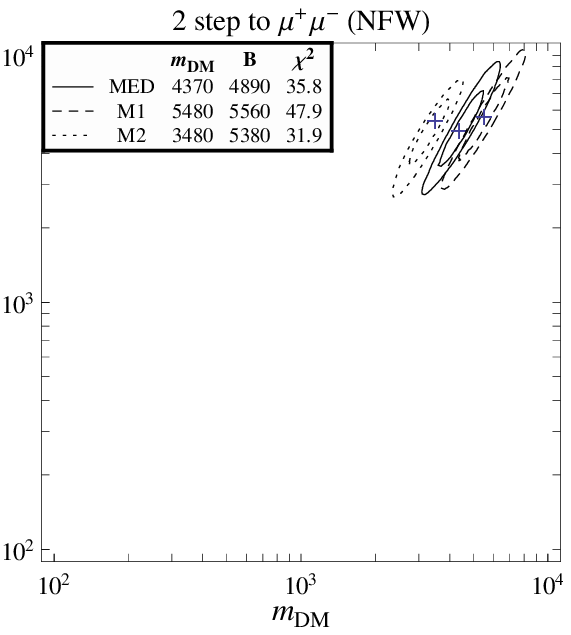}
  \includegraphics[scale=0.98]{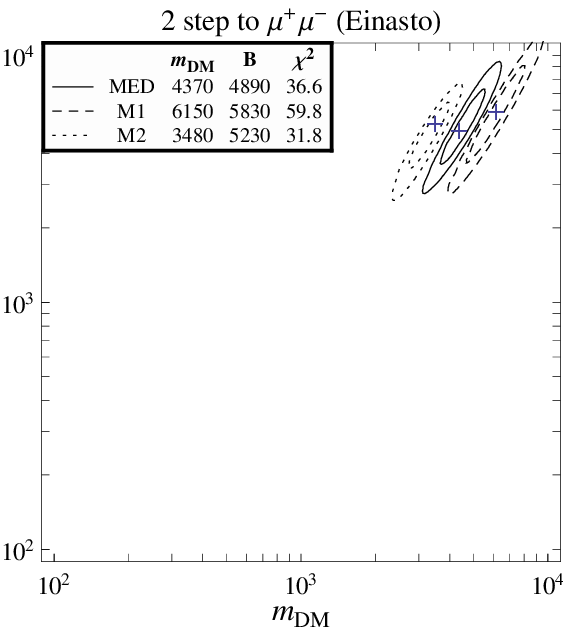}}
\caption{The same as Figure~\ref{fig:positron_e} but for annihilations 
 into $\mu^+ \mu^-$.}
\label{fig:positron_mu}
\end{figure}
The results of the fit for direct, $1$-step, and $2$-step 
annihilations are shown in the case of electron final states 
in Figure~\ref{fig:positron_e} and muon final states in 
Figure~\ref{fig:positron_mu}.  (The results do not depend on 
whether the dark matter particle is self-conjugate or not.) 
Each plot corresponds to a definite cascade annihilation pattern 
with a definite halo profile, and shows $1\sigma$ and $2\sigma$ 
contours for three propagation models in the $m_{\rm DM}$-$B$ plane. 
The best fit values for $m_{\rm DM}$ and $B$, as well as the $\chi^2$ 
values, are also given, where $\chi^2$ follows the chi-squared 
distribution with $22$ degrees of freedom ($7~\mbox{PAMELA} + 
21~\mbox{ATIC} - 6~\mbox{fitting parameters: }m_{\rm DM},B,A_\pm,P_\pm$). 
To evaluate goodness of fit, we also give $p$-values in Table~\ref{tab:1-p}, 
where we have chosen the propagation model giving smallest $\chi^2$ 
for each plot of Figures~\ref{fig:positron_e} and \ref{fig:positron_mu}. 
As we can see, the fits are reasonable for all the cases presented.
\begin{table}
\begin{center}
\begin{tabular}{c|ccc|ccc}
  & Direct $e$ & 1 step $e$ & 2 step $e$ & 
    Direct $\mu$ & 1 step $\mu$ & 2 step $\mu$ 
\\ \hline
  Isothermal &  $16\%$ & $17\%$ & $9.8\%$ & $12\%$ & $8.5\%$ & $7.7\%$ \\
  NFW        &  $13\%$ & $17\%$ & $9.4\%$ & $12\%$ & $7.7\%$ & $7.9\%$ \\
  Einasto    & $9.0\%$ & $18\%$ & $8.8\%$ & $12\%$ & $7.1\%$ & $8.1\%$ 
\end{tabular}
\end{center}
\caption{The $p$-value for the best propagation model for each plot in 
 Figures~\ref{fig:positron_e} and \ref{fig:positron_mu}.}
\label{tab:1-p}
\end{table}
\begin{figure}
  \center{\includegraphics[scale=0.65]{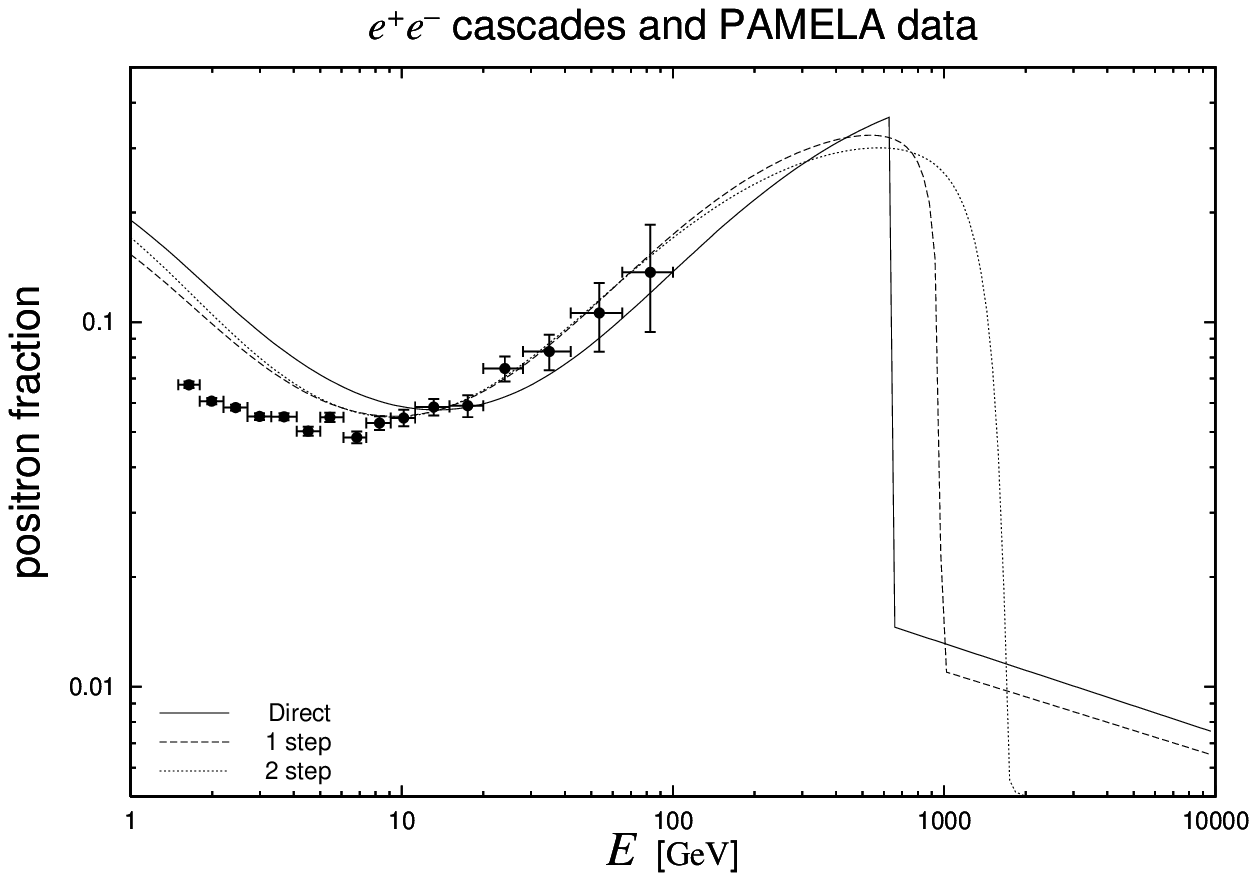}
  \hspace{0.5cm}
  \includegraphics[scale=0.65]{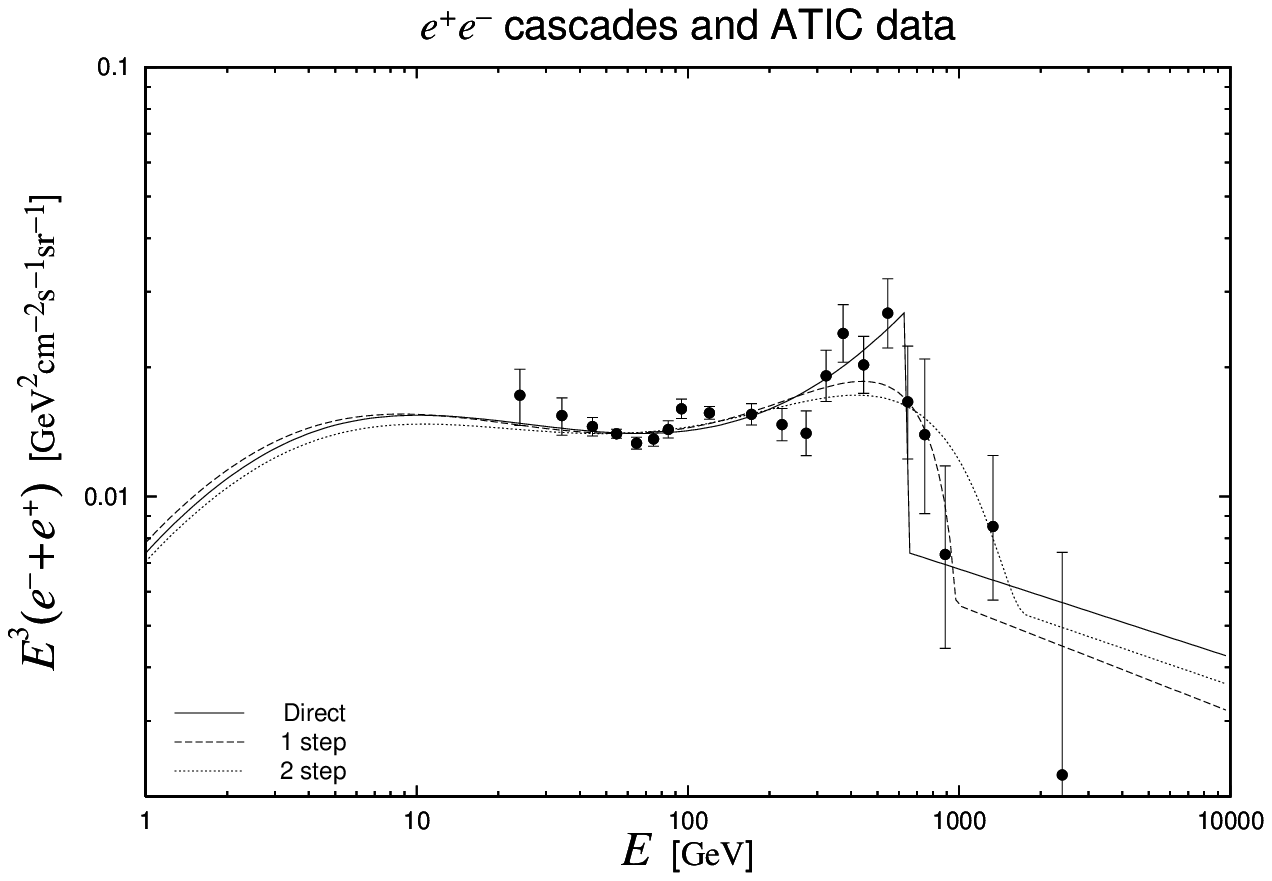}}
\caption{The predicted $e^\pm$ intensities compared to the PAMELA (left) 
 and ATIC (right) data for direct (solid), $1$-step (dashed), and $2$-step 
 (dotted) annihilations into electron final states.  The NFW halo profile 
 and the MED propagation model are chosen, and the $e^\pm$ backgrounds 
 are marginalized as described in Eq.~(\ref{eq:marginalize}).  Note 
 that we fit the PAMELA data only for $E \simgt 10~{\rm GeV}$ because 
 solar modulation effects are important at lower energies.}
\label{fig:spectra_e}
\end{figure}
\begin{figure}
  \center{\includegraphics[scale=0.65]{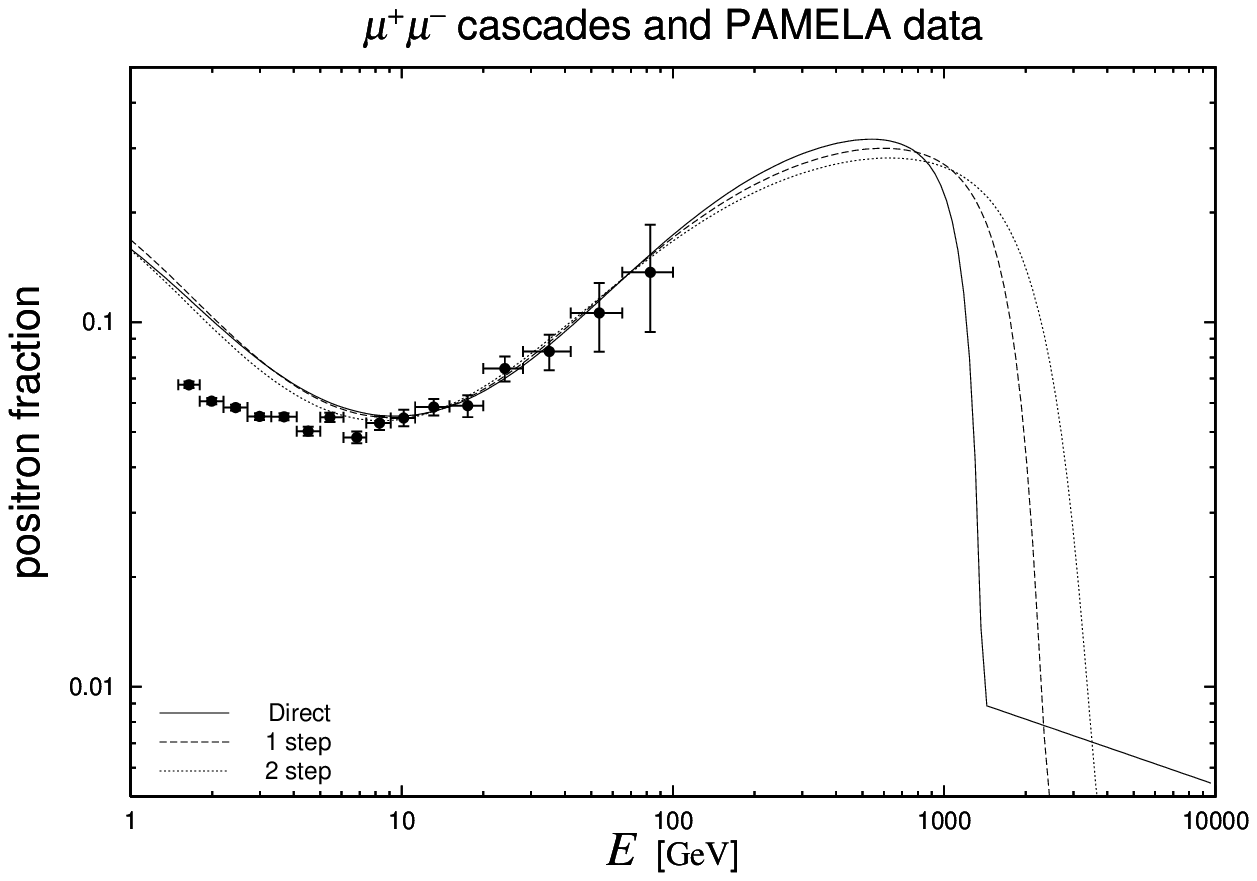}
  \hspace{0.5cm}
  \includegraphics[scale=0.65]{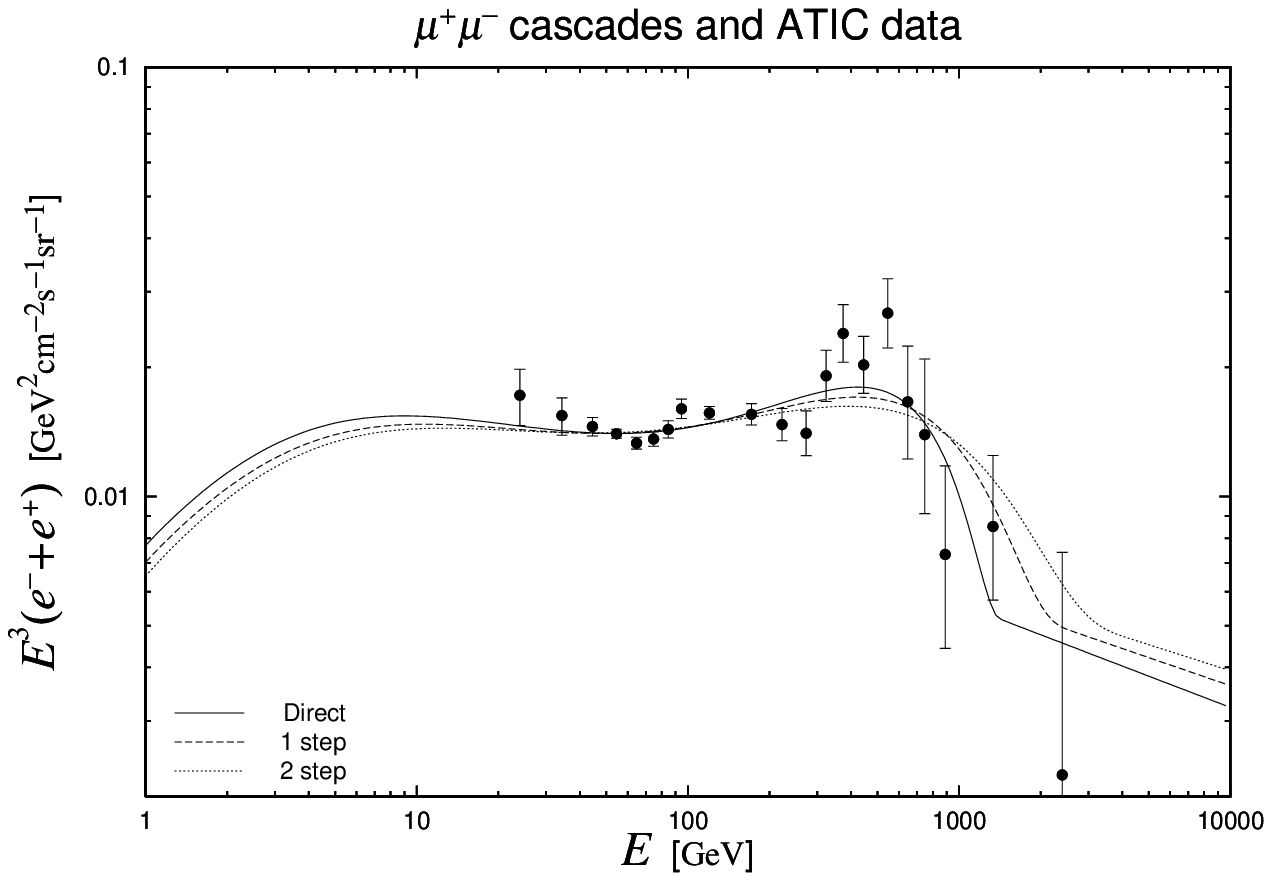}}
\caption{The same as Figure~\ref{fig:spectra_e} but for annihilations 
 into muon final states.}
\label{fig:spectra_mu}
\end{figure}
In Figures~\ref{fig:spectra_e} and \ref{fig:spectra_mu}, we show the 
comparisons of the $e^\pm$ intensities with the PAMELA and ATIC data 
in the cases of electron and muon final states.  Solid, dashed, and 
dotted lines represent direct, $1$-step, and $2$-step annihilations, 
and we have chosen the NFW halo profile and the MED propagation 
model for illustrative purposes.  We clearly see the trend that more 
steps in cascades lead to flatter $e^\pm$ spectra, but that the fits 
are good for all cases shown.  Note that we have not optimized the 
propagation model here.  Adjusting the propagation model can lead 
to a better fit in certain cases, especially the $2$-step annihilation 
into $\mu^+\mu^-$.

Summarizing the analysis of the $e^\pm$ data, we find:
\begin{itemize}
\item The PAMELA and ATIC data require the dark matter mass 
and boost factor in the region $m_{\rm DM} = O({\rm TeV})$ and 
$B = O(1000)$, which is consistent with earlier analyses on the 
direct~\cite{Cirelli:2008pk} and $1$-step~\cite{Cholis:2008wq} cases. 
More steps in the cascade lead to larger values of $m_{\rm DM}$ and 
$B$, and roughly speaking, both $B$ and $m_{\rm DM}$ scale as $2^n$. 
The reason is that the peak location in the ATIC data sets the scale 
for the final state $e^\pm$ energy, and since the average $e^\pm$ 
energy over $m_{\rm DM}$ in an $n$-step cascade scales like $1/2^n$, 
$m_{\rm DM}$ must increase by $2^n$ to keep the peak location fixed. 
Similarly, the annihilation signal scales like $B N_e/m_{\rm DM}^2$ 
(assuming fixed $\rho_{\rm DM}$), where $N_e = 2^{n+1}$ is the final 
state $e^\pm$ multiplicity.  Thus, to keep the PAMELA/ATIC rate 
fixed, $B$ must scale like $2^n$.
\item The fits do not become much worse by going to multiple steps, 
due to uncertainties in the highest energy ATIC data and uncertainties 
in the $e^\pm$ propagation model.  In particular, $2$-step annihilations 
still fit the data reasonably well for both electron and muon final 
states.  The required boost factors are rather large in the case of 
muon final states: $B$ of a few thousand.  Such large boost factors 
may come from both astrophysics, e.g.\ uncertainties in $\rho_\odot$ 
and nearby clumps of dark matter, and particle physics.
\item Uncertainties in halo profiles and propagation models do not 
significantly affect the dark matter mass and the boost factor. 
Errors from these uncertainties are mostly of $O(10\%)$ and at most 
a factor of $2$.  The $2\sigma$ ranges of the fits then determine 
$m_{\rm DM}$ and $B$ up to a factor of a few.
\end{itemize}

Recent measurements at H.E.S.S.~\cite{HESS} of the electron plus positron 
flux above $600~{\rm GeV}$~\cite{Aharonian:2008aaa} are qualitatively 
consistent with the ATIC spectrum.  Given the large systematic energy 
uncertainties and hadronic background, we do not use the H.E.S.S. data 
in our fits, although we remark that the observed steepening of the 
spectrum places some bounds on very long cascade decays and may disfavor 
spectra with a hard cutoff such as direct $e^\pm$ annihilation.

\section{Gamma Ray Constraints}
\label{sec:gamma}

When dark matter annihilates into charged leptons, there is a primary 
source of gamma rays coming from FSR.  Various gamma ray telescopes 
have looked for excess gamma rays coming from the galactic center, 
and the null result of such searches puts bounds on dark matter 
annihilation into charged leptons.  An additional effect that is 
beyond the scope of this paper is ICS, where electrons/positrons 
from dark matter annihilation lose energy by upscattering starlight 
photons into gamma rays.  The rate of ICS photon production depends 
on the modeling of galactic starlight, and we here focus only on 
the bounds from FSR.  For an early analysis of FSR in dark matter 
annihilation, see~\cite{Birkedal:2005ep}.

There is negligible energy loss as gamma rays propagate from the 
galactic center to the earth.  The total power (flux per energy) 
on earth depends on the dark matter halo profile through
\begin{equation}
  \frac{d\Phi_\gamma}{dE} 
  = \frac{B_{\gamma,{\rm astro}}}{8\pi\eta\, m_{\rm DM}^2} 
    \langle \sigma v \rangle \bar{J} \Delta\Omega\, \frac{d N_\gamma}{dE},
\label{eq:photon-flux}
\end{equation}
where $B_{\gamma,{\rm astro}}$ is an astrophysical boost factor for 
photons that may differ from $B_{e,{\rm astro}}$, and $dN_\gamma/dE$ 
is the photon energy spectrum per dark matter annihilation.  The energy 
spectrum from FSR is reviewed in Appendix~\ref{subapp:FSR}, and we 
also include the effect of radiative muon decays as described in 
Appendix~\ref{subapp:gamma-mu}.  $\Delta \Omega$ is the solid angle 
integration region, and $\bar{J}$ is the average line-of-sight-integrated 
squared dark matter density for a given halo model
\begin{equation}
  \bar{J} = \frac{1}{\Delta\Omega} \int_{\Delta\Omega}\! d\Omega 
    \int_{\rm line-of-sight}\! ds\, \rho(\vec{x})^2.
\label{eq:Jbar}
\end{equation}

The strongest bounds on FSR gamma rays come from atmospheric Cerenkov 
telescopes, but the way these experiments extract gamma ray signals 
affects the final dark matter annihilation bounds.  To enable background 
subtraction, these telescopes operate either in on-off mode or wobble 
mode, meaning the effective $\bar{J}$ exposure is~\cite{Mack:2008wu}
\begin{equation}
  \bar{J}_{\rm eff} = \bar{J}_{\rm on-source} - \bar{J}_{\rm off-source}.
\label{eq:Jbar-subt}
\end{equation}
By definition, $\bar{J}_{\rm eff}$ is smaller than 
$\bar{J}_{\rm on-source}$, and neglecting the $\bar{J}_{\rm off-source}$ 
contribution gives bounds that are too aggressive.  This means that 
one must know both the on- and off-source integration regions to derive 
a bound on the cross section.  For shallow dark matter halo profiles, 
there can be large cancellations in the value of $\bar{J}_{\rm eff}$, 
and in principle a stronger bound could be obtained using the raw 
unsubtracted data.

\begin{table}
\begin{center}
\begin{tabular}{c|ccc|ccc}
  & GC (on) & GC (off) & GC (eff) & GR (on) & GR (off) & GR (eff) 
\\ \hline
  Isothermal &             $10$ &             $10$ &          $0.028$ & 
                           $10$ &             $10$ &          $0.019$ \\
  NFW        & $1.1 \cdot 10^4$ & $3.6 \cdot 10^2$ & $1.1 \cdot 10^4$ & 
               $1.8 \cdot 10^3$ & $4.3 \cdot 10^2$ & $1.4 \cdot 10^3$ \\
  Einasto    & $5.8 \cdot 10^3$ & $7.3 \cdot 10^2$ & $5.1 \cdot 10^3$ & 
               $2.3 \cdot 10^3$ & $8.7 \cdot 10^2$ & $1.5 \cdot 10^3$ 
\end{tabular}
\end{center}
\caption{$\bar{J}$ values for GC and GR gamma ray observations (on-source, 
 off-source, and effective) in units of ${\rm GeV}^2\, {\rm cm}^{-6}\, 
 {\rm kpc}$.}
\label{tab:Jbar-GC-GR}
\end{table}
We set bounds on FSR using three H.E.S.S. gamma ray data sets.  The first 
two are observations of the Galactic Center (GC)~\cite{Aharonian:2006wh} 
and the Galactic Ridge (GR)~\cite{Aharonian:2006au}.  Neither is ideal 
for dark matter observations because of the large contamination from 
gamma ray point sources and molecular gas, and in principle one should 
put bounds on a dark matter signal after subtracting both these foregrounds. 
Since such subtractions are not available, we derive conservative 
bounds by insisting that the dark matter signal does not exceed any 
of the H.E.S.S. data points by more than $2\sigma$.  For the GC and 
GR samples, values of $\bar{J}$ for the three dark matter halos in 
Eqs.~(\ref{eq:rho-Isothermal},~\ref{eq:rho-NFW},~\ref{eq:rho-Einasto}) 
are shown in Table~\ref{tab:Jbar-GC-GR} for both the on-source and 
off-source regions.

The GC data set comes from the inner $0.1^\circ$ of the galaxy with 
a solid angle integration of $\Delta\Omega = 1 \times 10^{-5}$, 
corresponding to the gamma ray source HESS~J1745-290.  This sample 
was taken in wobble mode, and the off-source region corresponds to 
a ring at a distance of $1.4^\circ$ from the GC.  Apart from the 
off-source subtraction, no other corrections were made to the data, 
so the data points include both the HESS~J1745-290 point source 
as well as any putative dark matter signal.

The GR data set comes from the region $|\ell| < 0.8^\circ$, 
$|b| < 0.3^\circ$ in galactic coordinates, with foreground point 
sources HESS~J1745-290 and G$0.9+0.1$ subtracted.  The GR sample was 
taken in on-off mode, and the region  $|\ell| < 0.8^\circ$, $0.8^\circ 
< |b| < 1.5^\circ$ was used for background subtraction.  In the GR data, 
H.E.S.S. finds that the gamma ray emissions are spatially correlated 
with molecular gas traced by CS emission lines, but a molecular gas 
foreground component is not subtracted from the data.  The GR bounds 
are expected to strengthen if one were to subtract a molecular 
gas component.

H.E.S.S. looked more directly for dark matter annihilation in the 
Sagittarius dwarf spheroidal galaxy (Sgr~dSph)~\cite{Aharonian:2007km}. 
Sgr~dSph has negligible foregrounds and is thought to be dark matter 
dominated.  A $95\%$~C.L. model-independent bound on the gamma ray 
flux was obtained for $E_\gamma > 250~{\rm GeV}$:
\begin{equation}
  \Phi_\gamma(E_\gamma > 250~{\rm GeV}) 
  < 3.6 \times 10^{-12}~{\rm cm}^{-2}\, {\rm s}^{-1},
\label{eq:HESS-Sgr}
\end{equation}
with a solid angle integration of $\Delta\Omega = 2 \times 10^{-5}$. 
Since the Sgr~dSph data was taken in wobble mode and the contribution 
from the Sgr~dSph halo is negligible in a $1.4^\circ$ ring, we 
use $\bar{J}_{\rm eff} = \bar{J}_{\rm on-source}$.  The value of 
$\bar{J}_{\rm eff}$ strongly depends on the halo profile of Sgr~dSph. 
For example, an NFW profile, a large core profile, and a small core 
profile quoted in~\cite{Bertone:2008xr} lead to
\begin{equation}
  \bar{J}_{\rm NFW}^{\rm Sgr\,dSph} = 7.8 \times 10^2,
\qquad
  \bar{J}_{\rm Large\,Core}^{\rm Sgr\,dSph} = 1.1 \times 10^2,
\qquad
  \bar{J}_{\rm Small\,Core}^{\rm Sgr\,dSph} = 2.4 \times 10^4,
\label{eq:Jbar-Sgr}
\end{equation}
in units of ${\rm GeV}^2\, {\rm cm}^{-6}\, {\rm kpc}$.  To derive bounds, 
we consider NFW and large core profiles.

\begin{figure}
  \center{\includegraphics[scale=0.98]{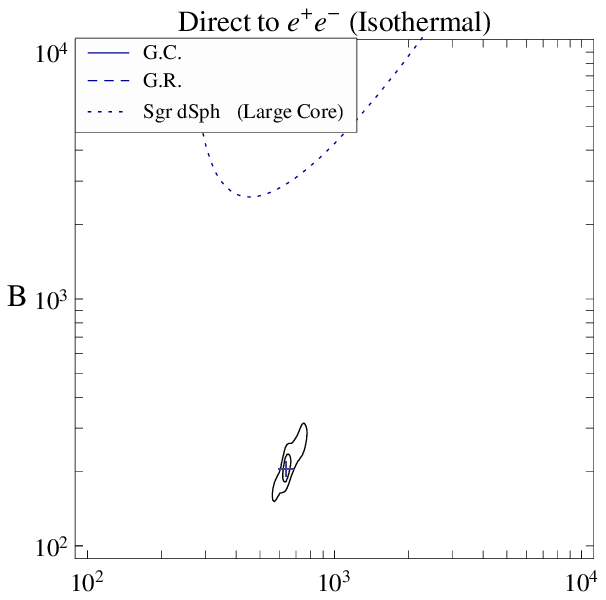}
  \includegraphics[scale=0.98]{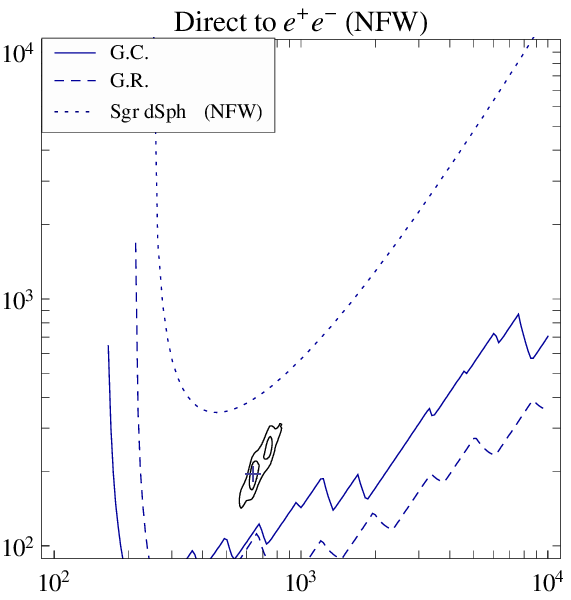}
  \includegraphics[scale=0.98]{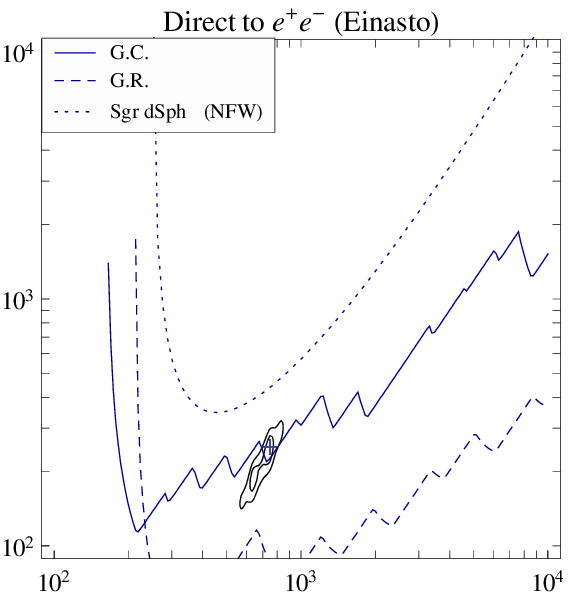}}
  \center{\includegraphics[scale=0.98]{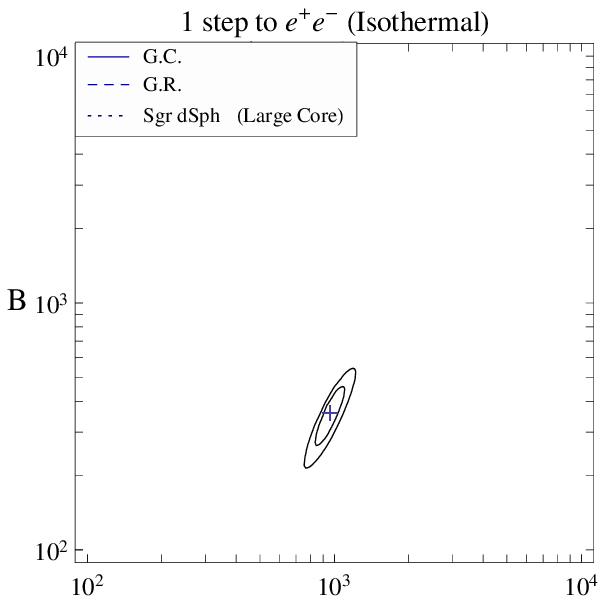}
  \includegraphics[scale=0.98]{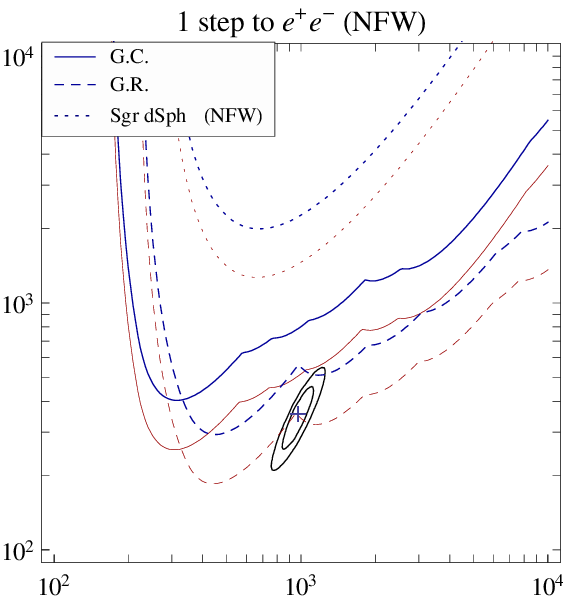}
  \includegraphics[scale=0.98]{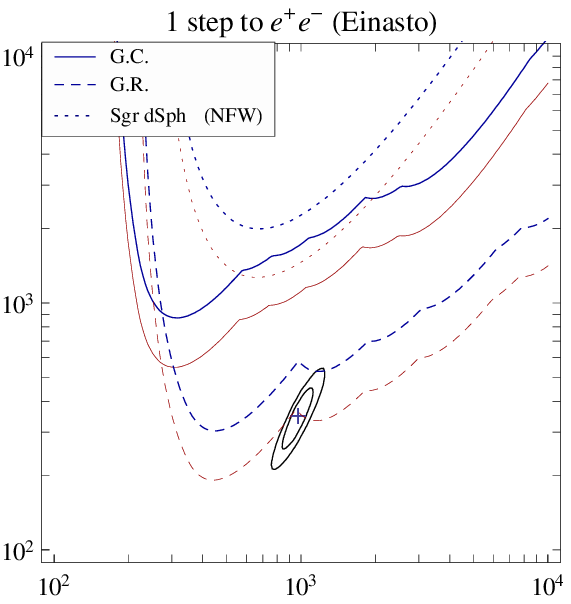}}
  \center{\includegraphics[scale=0.98]{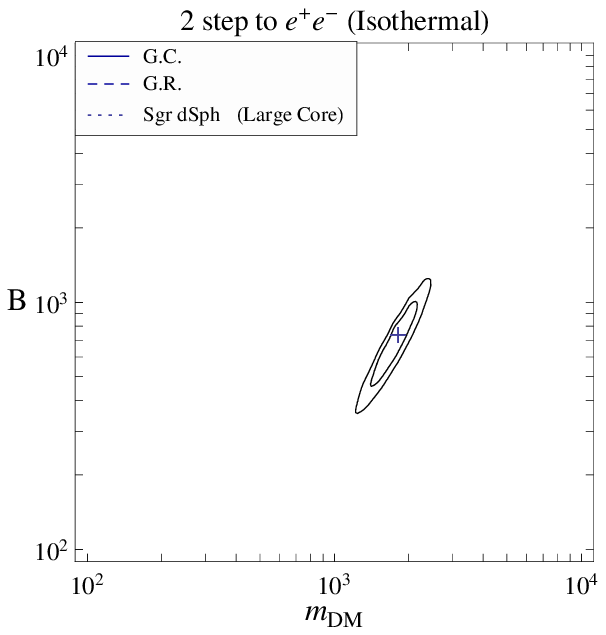}
  \includegraphics[scale=0.98]{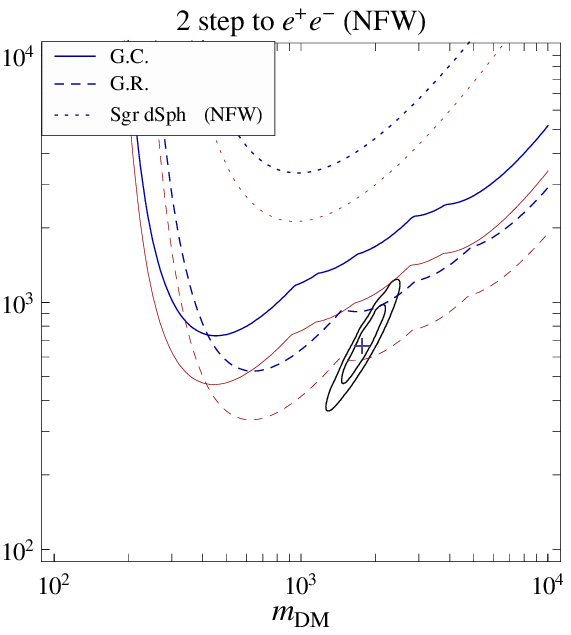}
  \includegraphics[scale=0.98]{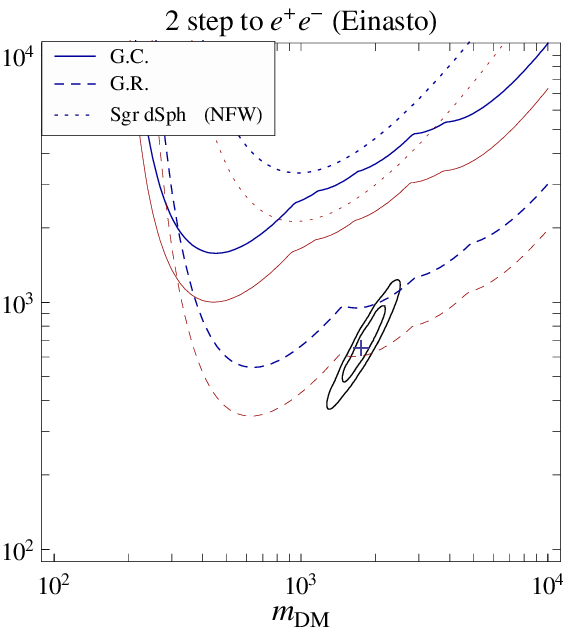}}
\caption{Constraints from gamma ray observations, GC (solid), GR (dashed), 
 and Sgr~dSph (dotted), in the $m_{\rm DM}$-$B$ plane for direct, 
 $1$-step, and $2$-step annihilations into electron final states. 
 All the constraints, as well as the best fit region for PAMELA/ATIC 
 (MED propagation), are plotted assuming $B_{e,{\rm astro}} = 
 B_{\gamma,{\rm astro}}$.  For cascade annihilations, each of the GC, 
 GR, and Sgr~dSph constraints consist of two curves, with the upper 
 (blue) and lower (red) curves corresponding to $m_1 = 100~{\rm MeV}$ 
 and $1~{\rm GeV}$, respectively.  Note that the constraint lines in 
 the cored isothermal case are above the plot region, and that the 
 halo profiles for Sgr~dSph are given in the legends.}
\label{fig:constraint_e}
\end{figure}
\begin{figure}
  \center{\includegraphics[scale=0.98]{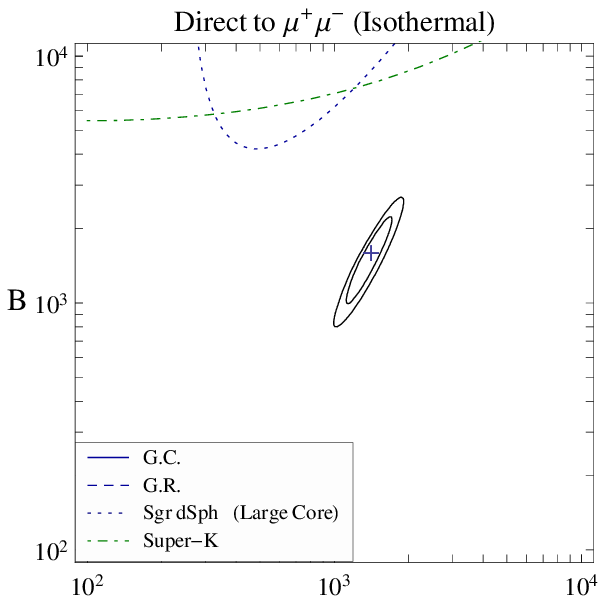}
  \includegraphics[scale=0.98]{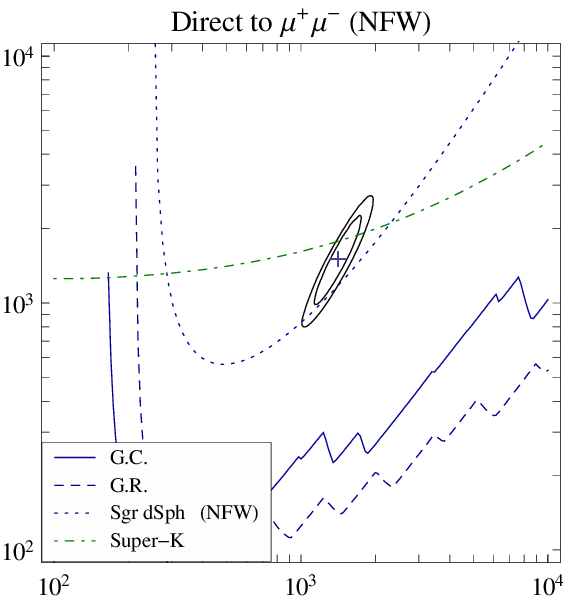}
  \includegraphics[scale=0.98]{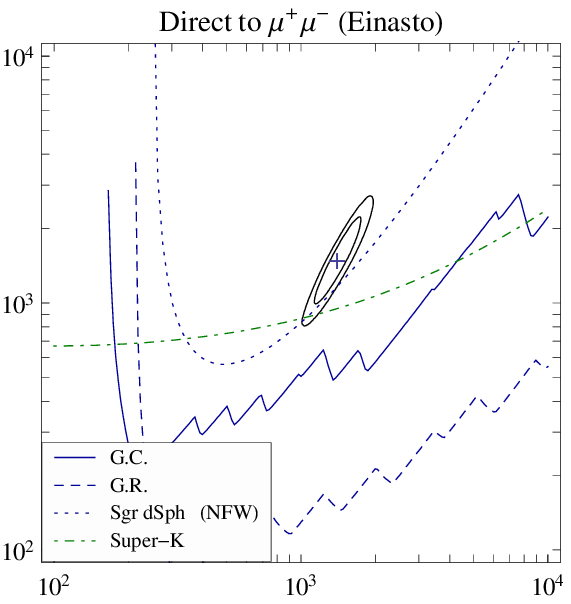}}
  \center{\includegraphics[scale=0.98]{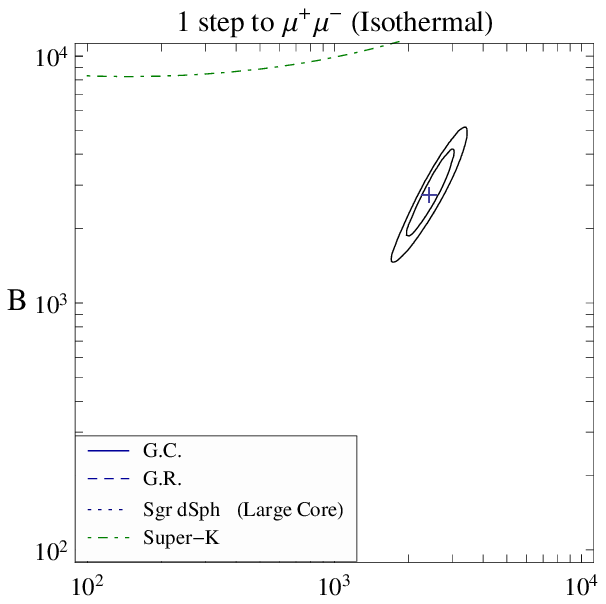}
  \includegraphics[scale=0.98]{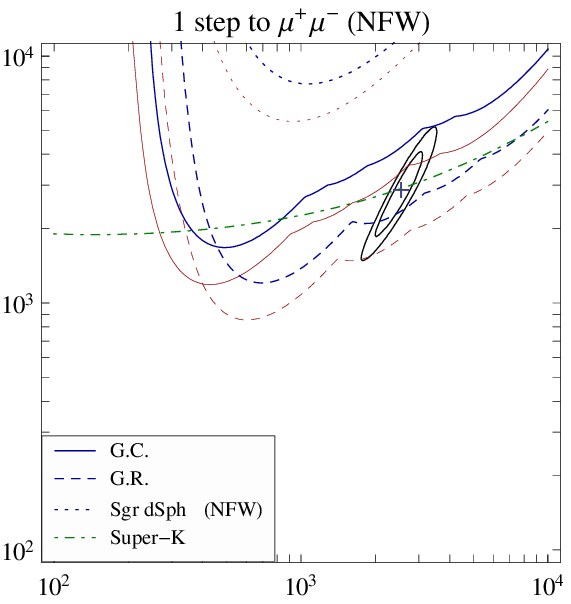}
  \includegraphics[scale=0.98]{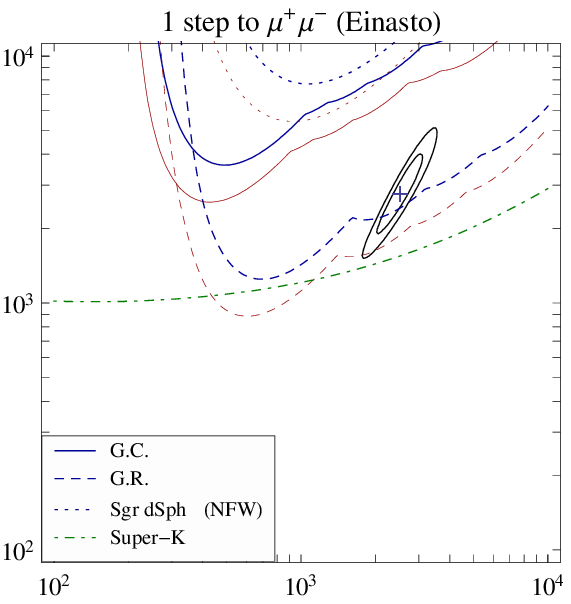}}
  \center{\includegraphics[scale=0.98]{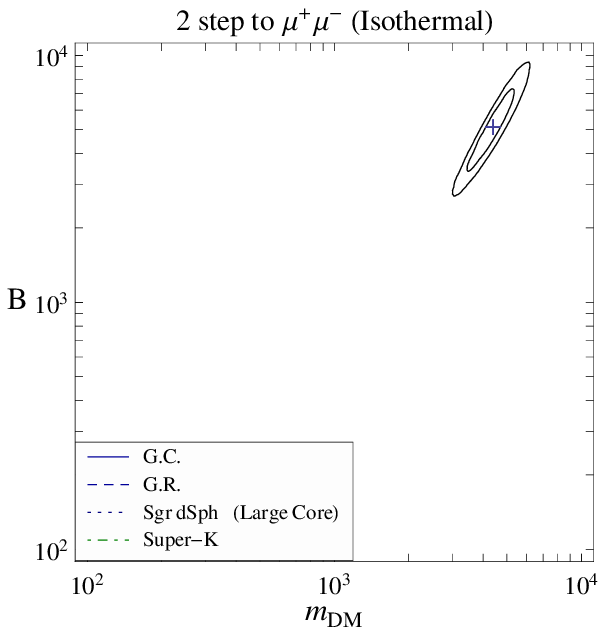}
  \includegraphics[scale=0.98]{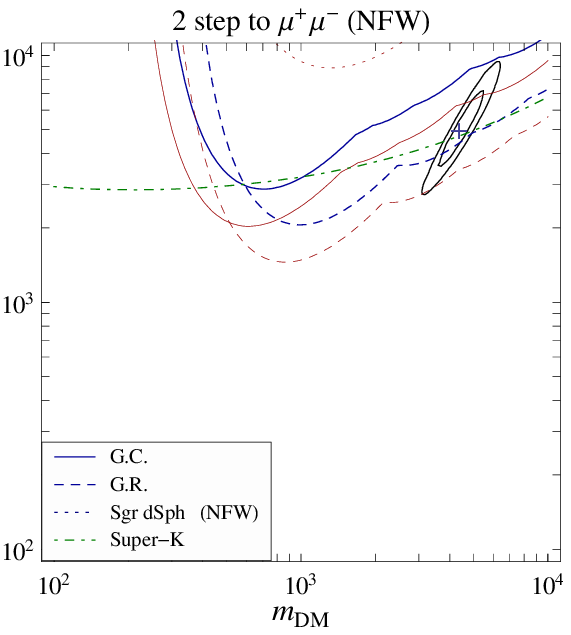}
  \includegraphics[scale=0.98]{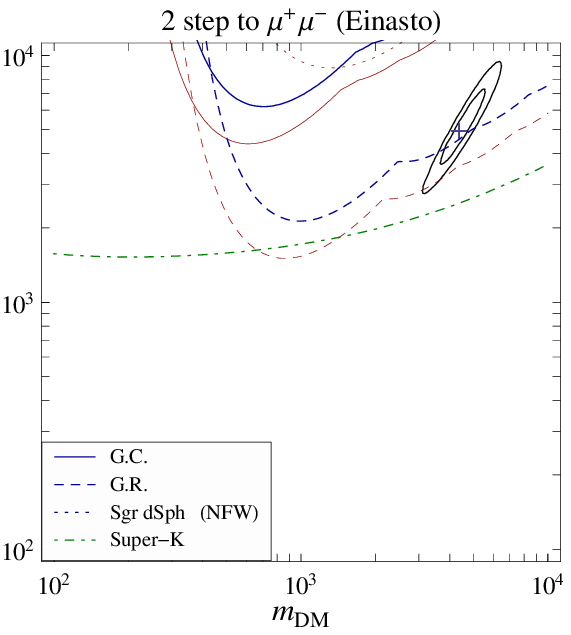}}
\caption{The same as Figure~\ref{fig:constraint_e} but for muon final 
 states.  Also included are constraints from neutrino observations 
 (dot-dashed) assuming $B_{\nu,{\rm astro}} = B_{e,{\rm astro}}$. 
 For cascade annihilations, the upper (blue) and lower (red) curves 
 now correspond to $m_1 = 600~{\rm MeV}$ and $1~{\rm GeV}$, respectively.}
\label{fig:constraint_mu}
\end{figure}
In Figures~\ref{fig:constraint_e} and \ref{fig:constraint_mu}, we show the 
resulting constraints from the GC, GR, and Sgr~dSph gamma ray observations 
in the $m_{\rm DM}$-$B$ plane.  We also superimpose the $1\sigma$ and 
$2\sigma$ contours reproducing the PAMELA/ATIC data for the MED propagation 
model; see Figures~\ref{fig:positron_e} and \ref{fig:positron_mu}. 
In order to plot the Sgr~dSph and Milky Way bounds on the same plane, 
we associate the Sgr~dSph small core profile with the Milky Way cored 
isothermal profile, and the Sgr~dSph NFW profile with the Milky Way NFW 
and Einasto profiles.  The ragged lines in the GC and GR constraints 
come from the binning of the H.E.S.S. data.

We note that here we have drawn the gamma ray constraints and $e^\pm$ 
contours assuming a common $B_{\gamma,{\rm astro}}$ and $B_{e,{\rm astro}}$. 
This is most likely not the case.  For example, if dark matter 
clumping decreases toward the galactic center, then the gamma ray 
flux will decrease compared to the local positron flux.  If the 
astrophysical boost factor for $e^\pm$ is larger than that for 
$\gamma$, then the gamma ray constraints become weaker by a factor 
of $B_{e,{\rm astro}}/B_{\gamma,{\rm astro}}$ compared to those 
shown in Figures~\ref{fig:constraint_e} and \ref{fig:constraint_mu}. 

The bounds from Sgr~dSph can be modified if there are nonperturbative 
enhancements to the dark matter annihilation cross section.  Since 
the velocity dispersion of dark matter in Sgr~dSph is $v_{\rm Sgr} 
\sim 10~{\rm km/s}$~\cite{Evans:2003sc}, as opposed to $v_{\rm MW} 
\sim 200~{\rm km/s}$ in the Milky Way, the boost factor relevant for 
Sgr~dSph may be larger than that for electrons/positrons if part of 
the boost factor arises from the Sommerfeld or bound state enhancement. 
This would make the Sgr~dSph constraint stronger than what is naively 
read from Figures~\ref{fig:constraint_e} and \ref{fig:constraint_mu}.

Summarizing the analysis of the gamma ray constraints, we find:
\begin{itemize}
\item For FSR gamma rays from the galactic center region, the GR data 
gives somewhat stronger constraints than the GC data.  For direct 
annihilation, this disfavors the NFW and Einasto profiles, as also 
seen in Refs.~\cite{Bell:2008vx,Bertone:2008xr}.  The Sgr~dSph data 
is not constraining unless we were to take a highly peaked halo such 
as the small core profile.
\item The constraints from FSR photons are weaker in the cascade 
annihilation case than in the direct annihilation case.  This is because 
cascade annihilations give smaller photon yield at high energies, as 
discussed in Appendix~\ref{subapp:FSR}.  The smaller the $\phi_1$ mass 
is, the weaker the constraints become.  The constraints, however, do 
not become weaker by increasing the number of steps, as can be seen 
by comparing the best fit region for the $e^\pm$ data with the gamma 
ray constraints.  This is because the reduction in the photon yield 
depends only on the $\phi_1$ mass and not on the number of cascade 
steps, and the softening of the gamma ray spectra is compensated by 
the increase in the best fit dark matter mass and boost factor.
\item The constraints are very weak for shallow halo profiles 
such as the cored isothermal profile for the Milky Way in 
Eq.~(\ref{eq:rho-Isothermal}).  This is particularly true for 
the GC and GR data because the background subtraction due to on-off 
or wobble mode operation also subtracts (most of) the signals from 
dark matter annihilations.  The constraints from these data, therefore, 
are rather weak as long as the halo profile is relatively flat within 
about $100~{\rm pc}$ of the galactic center.  A better bound may 
be obtained if we could use the unsubtracted data.
\end{itemize}

While we have focused only on FSR photons in our analysis, we wish to 
make a few comments about ICS, the WMAP Haze, and radio bounds.  In the 
context of dark matter, the WMAP Haze arises because electrons from dark 
matter annihilation emit synchrotron radiation in the galactic magnetic 
fields.  The total synchrotron power---and hence the size of the 
predicted WMAP Haze signal---depends on whether these electrons can 
lose energy via non-synchrotron channels.  Ref.~\cite{Cholis:2008wq} 
found that in order to be consistent with the large boost factors 
necessary to explain PAMELA/ATIC, one had to assume a larger rate 
for ICS compared to earlier Haze analyses~\cite{Finkbeiner:2004us}. 
Given the uncertainty in galactic starlight and the dark matter halo 
profile, it is consistent to conservatively ignore potential bounds 
from ICS, but since the WMAP Haze is one of the motivations for 
considering dark matter annihilation, strictly speaking one should 
verify that the assumed electron energy loss mechanisms can yield 
the WMAP Haze while satisfying ICS photon bounds.

That said, we do not expect much variation in the ICS bounds 
between direct and cascade annihilation scenarios.  ICS is calculated 
from a steady state configuration of charged particles, so to the 
extent that the PAMELA/ATIC data already normalizes the steady state 
electron/positron densities, the ICS yield should be similar regardless 
of the annihilation scenario.  Therefore, the recent analysis of 
Ref.~\cite{Cholis:2008wq} should be representative of generic multi-step 
cascade annihilation scenarios.  This is similar in spirit to the 
WMAP Haze, in that the Haze requires a source of charged particles 
to generate the synchrotron signal, but the precise particle energy 
distribution has only a secondary effect.

Finally, there has also been recent 
suggestions~\cite{Bertone:2008xr,Bergstrom:2008ag} of 
a possible tension between a dark matter annihilation interpretation 
of the WMAP Haze and bounds from $408~{\rm MHz}$ radio observations in 
the inner $4''$ of the Milky Way~\cite{Davies:1976}.  Such bounds assume 
that the steep halo profiles necessary to generate the Haze at a latitudinal 
distance between $5^\circ$ and $30^\circ$ can be extrapolated to sub-parsec 
distances away from the galactic center.  There are a number of reasons 
to distrust such an extreme extrapolation of the dark matter halo, 
including possible effects of baryons~\cite{Dutton:2006vi} and hierarchical 
mergers~\cite{Merritt:2002vj} to soften cuspy behavior.  At minimum, 
$N$-body simulations~\cite{Diemand:2008in} do not have resolution to 
such small scales.  Therefore, we find no reason to disfavor a dark 
matter annihilation scenario on the basis of the $408~{\rm MHz}$ 
radio bound.  Note that the analysis of Ref.~\cite{Borriello:2008gy} 
using an all-sky radio model~\cite{deOliveiraCosta:2008pb} finds only 
relatively mild synchrotron constraints on TeV-scale dark matter 
after masking the inner $15^\circ \times 15^\circ$ of the galaxy.

\section{Neutrino Constraints}
\label{sec:neutrino}

When dark matter annihilates into muons, there is an irreducible source 
of neutrinos.  Neutrinos produced in the galactic center oscillate as 
they travel towards earth, and if they are muon-type neutrinos when 
they collide with rock in the earth's crust, they can create an 
upward-going flux of muons.  These muons could be observed by water 
Cerenkov detectors, and the absence of such observations puts bounds 
on the dark matter annihilation rate into muons.  For muon cascades, 
there is no high energy neutrino source from dark matter that accretes 
in the sun and earth, because the muons from dark matter annihilation 
are stopped before they decay~\cite{Ritz:1987mh}.

Since neutrinos have negligible energy losses as they traverse the 
galaxy, the muon-neutrino flux incident on earth is
\begin{equation}
  \frac{d\Phi_{\nu_\mu}}{dE_\nu} = \sum_i  P_{\nu_i \nu_\mu} 
    \frac{B_{\nu,{\rm astro}}}{8\pi\eta\, m_{\rm DM}^2} 
    \langle \sigma v \rangle \bar{J} \Delta\Omega\, \frac{d N_{\nu_i}}{dE},
\label{eq:flux-nu}
\end{equation}
where $i$ runs over the neutrino flavors, $B_{\nu,{\rm astro}}$ is the 
astrophysical boost factor for the neutrino signal which could differ 
from $B_{e,{\rm astro}}$, and $P_{\nu_i \nu_\mu}$ is the probability 
that $\nu_i$ has oscillated into $\nu_\mu$~\cite{Amsler:2008zzb}:
\begin{equation}
  P_{\nu_\mu \nu_\mu} = 0.39,
\qquad
  P_{\nu_e \nu_\mu} = 0.21.
\label{eq:P-nu-osc}
\end{equation}
There is an analogous formula for $\bar{\nu}_\mu$, but in dark matter 
annihilations the $\nu_\mu$ and $\bar{\nu}_\mu$ fluxes are equal.  The 
primary neutrino spectra are given in Appendix~\ref{subapp:muon-decay}.

We now calculate the resulting upward-going muon flux following the 
analysis of Ref.~\cite{Barger:2007xf}.  An incident neutrino of energy 
$E_\nu$ creates muons of energy $E_\mu$ according to the neutrino-nucleon 
scattering cross sections $\sigma_{\nu N \rightarrow \mu X}$.  For 
the propagation of created muons, we use an approximate energy loss 
parameterization
\begin{equation}
  \frac{dE}{dL} = \rho_{\rm mat}(-\alpha - \beta E),
\label{eq:E-loss}
\end{equation}
with ``standard rock'' values $\alpha = 2 \times 10^{-6}~{\rm TeV}\, 
{\rm cm}^2/{\rm g}$ and $\beta = 4 \times 10^{-6}~{\rm cm}^2/{\rm g}$; 
$\rho_{\rm mat}$ will cancel in the final muon flux expression. 
In this approximation, a muon of starting energy $E_\mu$ can travel 
a distance
\begin{equation}
  L(E_\mu, E_{\rm thres.}) = \frac{1}{\rho_{\rm mat} \beta} \ln\left( 
    \frac{\alpha + \beta E_\mu}{\alpha + \beta E_{\rm thres.}} \right)
\label{eq:travel-dist}
\end{equation}
before its energy drops below the muon detection threshold $E_{\rm thres.}$.

For a given $d\Phi_{\nu_\mu}/dE_\nu$, the observed muon flux is
\begin{equation}
  \Phi_{\mu} = \int_{E_{\rm thres.}}^{m_{\rm DM}}\! d E_\mu 
    \int_{E_\mu}^{m_{\rm DM}}\! d E_\nu\, n_N 
    \left( \frac{d\Phi_{\nu_\mu}}{dE_\nu} 
      \frac{d \sigma_{\nu_\mu N \rightarrow \mu^- X}}{d E_\mu} 
    + \frac{d\Phi_{\bar{\nu}_\mu}}{dE_\nu} 
      \frac{d \sigma_{\bar{\nu}_\mu N \rightarrow \mu^+ X}}{d E_\mu} 
    \right) L(E_\mu, E_{\rm thres.}),
\label{eq:muon-flux}
\end{equation}
where $n_N = \rho_{\rm mat}/m_N$ is the nucleon number density in the 
earth's crust, and $N$ refers to an average nucleon.  We calculate the 
neutrino-nucleon scattering cross sections assuming an equal fraction 
of protons and neutrons in rock, using CTEQ5M parton distribution 
functions~\cite{Lai:1999wy} to include the effect of sea quarks, and 
retaining the full $W$ boson propagator in the cross section.

\begin{table}
\begin{center}
\begin{tabular}{c|ccccccc}
  & $3^\circ$ & $5^\circ$ & $10^\circ$ & $15^\circ$ & 
    $20^\circ$ & $25^\circ$ & $30^\circ$ 
\\ \hline
  Isothermal &  $10$ &  $10$ & $9.7$ & $9.0$ & $8.2$ & $7.4$ & $6.6$ \\
  NFW        & $340$ & $190$ &  $84$ &  $51$ &  $35$ &  $26$ &  $20$ \\
  Einasto    & $640$ & $370$ & $160$ &  $91$ &  $60$ &  $43$ &  $32$ 
\\ \hline
  Super-K ($95\%$~C.L.) & 
    $2.70$ & $4.82$ & $6.43$ & $10.6$ & $11.2$ & $17.6$ & $19.5$ 
\end{tabular}
\end{center}
\caption{$\bar{J}$ values for neutrino observations in units of 
 ${\rm GeV}^2\, {\rm cm}^{-6}\, {\rm kpc}$, and Super-K $95\%$~C.L. 
 flux limits in units of $10^{-15}~{\rm cm}^{-2} \, {\rm s}^{-1}$.}
\label{tab:Jbar-nu}
\end{table}
Super-K~\cite{Fukuda:2002uc} placed $90\%$ confidence bounds 
on the upward-going muon flux~\cite{Desai:2004pq} in various 
cone sizes ranging from $3^\circ$ to $30^\circ$ around the 
galactic center, with $E_{\rm thres.} = 1.6~{\rm GeV}$. 
To be more conservative, we consider $95\%$ confidence 
bounds~\cite{Desai:2009}, as shown in Table~\ref{tab:Jbar-nu}. 
The relevant values of $\bar{J}$ for the three dark matter halos in 
Eqs.~(\ref{eq:rho-Isothermal},~\ref{eq:rho-NFW},~\ref{eq:rho-Einasto}) 
are also shown in Table~\ref{tab:Jbar-nu}.  To derive a bound on the 
annihilation rate, we insist that the predicted flux does not exceed 
the $95\%$ confidence bound for any of the Super-K cone sizes. 
The neutrino constraints for muon cascade scenarios appear in 
Figure~\ref{fig:constraint_mu}, assuming a common boost factor 
for the electron, gamma ray, and neutrino signals.

As observed in Ref.~\cite{Ritz:1987mh}, for sufficiently small dark 
matter masses, the observed muon flux is nearly independent of the dark 
matter mass.  The reason is that both the neutrino-nucleon scattering 
cross section and the muon range scale like energy, but the dark matter 
number density squared (and hence the annihilation signal) scales like 
$1/m_{\rm DM}^2$, so the final observed flux is simply related to the 
normalized second-moment of the neutrino energy spectrum.  Hence the 
exclusion limit for the neutrino boost factor is approximately flat 
in $m_{\rm DM}$ for small enough dark matter masses.  As the dark matter 
mass increases, the average neutrino gets harder, and the neutrino-nucleon 
cross section grows less steeply because of the $W$ boson propagator. 
In addition, the energy-dependent term in $dE/dL$ begins to take effect, 
relaxing the neutrino bounds for high dark matter masses.

Summarizing the analysis of the neutrino constraints, we find:
\begin{itemize}
\item Since the neutrinos are softer in cascade annihilations, the 
bounds on the boost factor are weaker than for direct annihilation. 
However, the electrons are also softer in cascade annihilations, so 
the PAMELA/ATIC best fit mass and boost factor rise.  Put together, 
the neutrino tension increases marginally as the number of cascade 
steps increase.  For direct annihilation, our results agree 
qualitatively with \cite{Hisano:2008ah,Liu:2008ci}.
\item Super-K considered solid angles as large as $30^\circ$, so 
a large fraction ($\simeq 30\%$ to $\simeq 70\%$) of the total dark 
matter annihilation signal is contained within the observed region. 
While the Einasto profile is less peaked than NFW toward the galactic 
center, the Einasto bound happens to be stronger because of the large 
integration region.  For the NFW and Einasto profiles, the dominant 
bounds come from the $10^\circ$ cone, while for the cored isothermal 
profile they come from the $30^\circ$ cone.
\item Since the gross structure of the dark matter halo is better 
understood than the halo density at the galactic center, the Super-K 
neutrino constraints are in some sense more robust than the H.E.S.S. 
gamma ray constraints.  As in the case of gamma rays, the bounds are 
rather weak for the cored isothermal profile.  They are, however, 
significantly stronger in the more realistic NFW and Einasto profiles. 
For cascade decays, the neutrino constraints are comparable to or 
stronger than the FSR gamma ray constraints, and highlight the tension 
in muon cascade scenarios.  As with the gamma ray bounds, however, 
differences between the astrophysical boost factors $B_{e,{\rm astro}}$ 
and $B_{\nu,{\rm astro}}$ could alleviate the tension.
\end{itemize}

\section{The Axion Portal}
\label{sec:axion}

\begin{figure}[t]
\begin{center}
\begin{picture}(130,90)(45,18)
  \Line(50,65)(80,50) \Text(49,65)[r]{$\chi$}
  \Line(50,35)(80,50) \Text(49,35)[r]{$\bar{\chi}$}
  \Line(80,50)(110,70) \Text(100,68)[br]{$s$}
  \Line(80,50)(110,30) \Text(100,32)[tr]{$a$}
  \GCirc(80,50){10}{0.7}
  \Line(110,70)(140,85) \Text(130,84)[br]{$a$}
  \Line(110,70)(140,60) \Text(130,59)[tr]{$a$}
  \Line(110,30)(140,40) \Text(155,35)[br]{$\ell^+$}
  \Line(110,30)(140,15) \Text(155,20)[tr]{$\ell^-$}
  \Line(140,85)(170,90) \Text(175,93)[l]{$\ell^+$}
  \Line(140,85)(170,80) \Text(175,81)[l]{$\ell^-$}
  \Line(140,60)(170,65) \Text(175,68)[l]{$\ell^+$}
  \Line(140,60)(170,55) \Text(175,56)[l]{$\ell^-$}
\end{picture}
\end{center}
\caption{In the axion portal, fermionic dark matter annihilates dominantly 
 into a scalar $s$ and a pseudoscalar ``axion'' $a$.  The scalar then 
 decays as $s \rightarrow aa$, and the axion decays as $a \rightarrow 
 \ell^+ \ell^-$.  In the minimal axion portal, the axion dominantly decays 
 into muons, but in the leptonic axion portal it can dominantly decay 
 into electrons.  These models are partway between a $1$-step and 
 a $2$-step cascade annihilation scenario.}
\label{fig:axion-portal}
\end{figure}
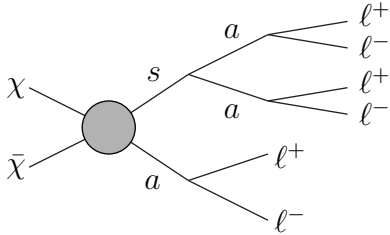
One well-motivated example of a cascade annihilation scenario is the 
axion portal~\cite{Nomura:2008ru}.   In this scenario, dark matter is 
a TeV-scale particle that obtains a mass from spontaneous symmetry 
breaking.  The spontaneous breaking of $U(1)_X$ yields a pseudoscalar 
``axion'' $a$ and a scalar ``Higgs'' $s$, and for fermionic dark 
matter the dominant annihilation channel is
\begin{equation}
  \chi \bar{\chi} \rightarrow s a
\label{eq:axion-annih}
\end{equation}
($\chi \chi \rightarrow s a$ if $\chi$ is a Majorana fermion).  The 
scalar $s$ dominantly decays as $s \rightarrow aa$, and if standard model 
leptons carry axial $U(1)_X$ charges, then $a \rightarrow \ell^+ \ell^-$. 
Since $a$ is a pseudoscalar, helicity suppression implies $a$ will 
decay into the heaviest kinematically allowed lepton, which we assume 
is either an electron or muon.  An exchange of $s$ can also provide 
the necessary enhancement of the annihilation cross section 
through the Sommerfeld and/or bound state effect.  As shown in 
Figure~\ref{fig:axion-portal}, the axion portal effectively gives 
a one-and-a-half step cascade annihilation in the language used here.

The simplest model for the axion portal---the minimal axion portal---is 
obtained if we identify $U(1)_X$ with a Peccei-Quinn symmetry rotating 
two Higgs doublets.  In this case, $a$ has large hadronic couplings, 
and there are strong constraints on the axion mass from beam dump 
experiments and rare meson decays.  Ref.~\cite{Nomura:2008ru} found 
that the preferred axion mass range was $360~{\rm MeV} \simlt m_a 
\simlt 800~{\rm MeV}$, in which case $a$ preferentially decays into 
muons.  In the analysis here, we also consider a variant of the axion 
portal---the leptonic axion portal---where only leptons are charged 
under $U(1)_X$.  This possibility was mentioned in~\cite{Nomura:2008ru}, 
and is described in more detail in Appendix~\ref{app:lepto-axion}. 
Dark matter in this model annihilates through the axion $a_\ell$ 
associated with the leptonic symmetry, which does not have a coupling 
to hadrons.  The absence of hadronic couplings allows the parameter 
range $2 m_e < m_{a_\ell} < 2 m_\mu$, so $a_\ell$ can preferentially 
decay into electrons.  Note that the nonperturbative enhancement 
of the halo cross section is caused by the exchange of another 
scalar $s_\ell$ and not by the axion $a_\ell$, so the bound of 
Ref.~\cite{Kamionkowski:2008gj} does not exclude $a_\ell$ masses 
smaller than $\approx 100~{\rm MeV}$.

\begin{figure}
  \center{\includegraphics[scale=0.98]{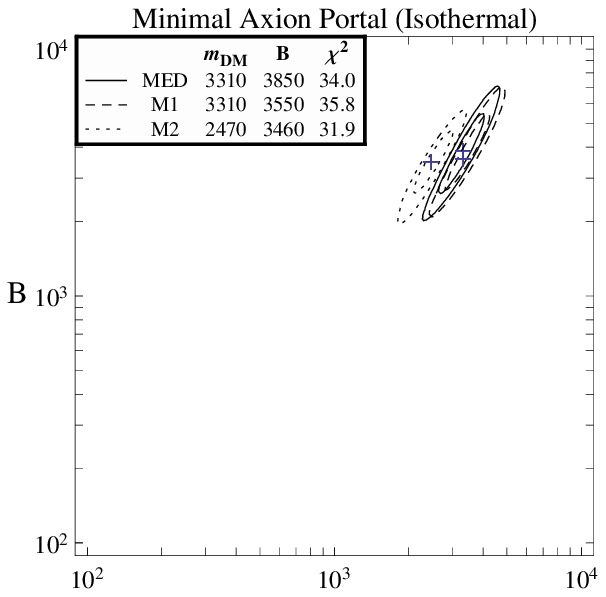}
  \includegraphics[scale=0.98]{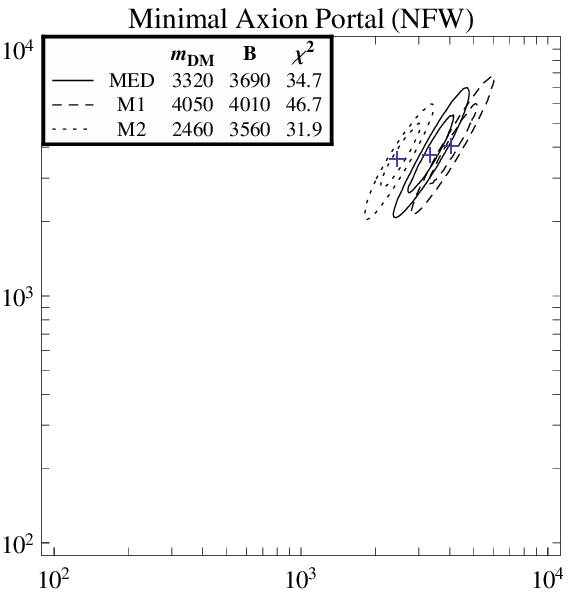}
  \includegraphics[scale=0.98]{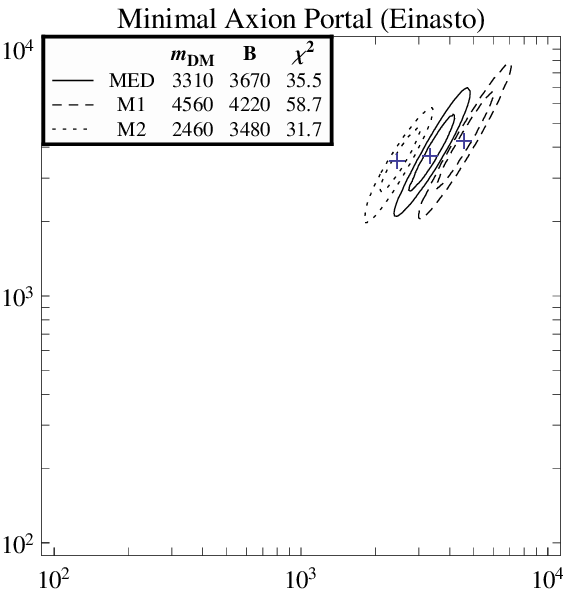}}
  \center{\includegraphics[scale=0.98]{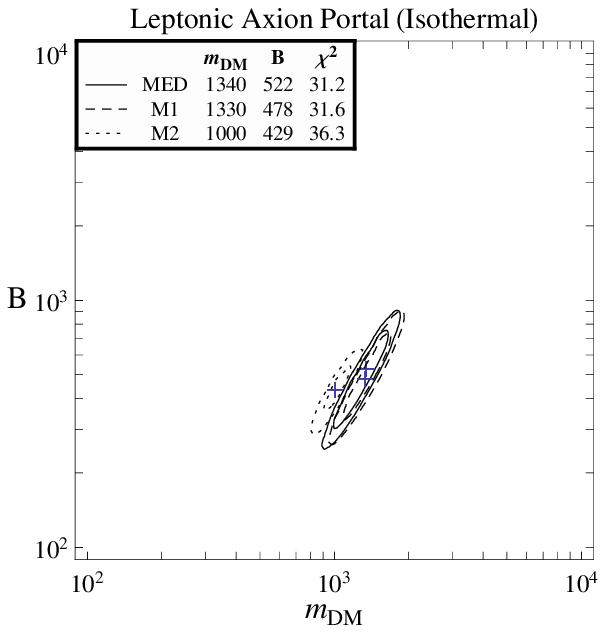}
  \includegraphics[scale=0.98]{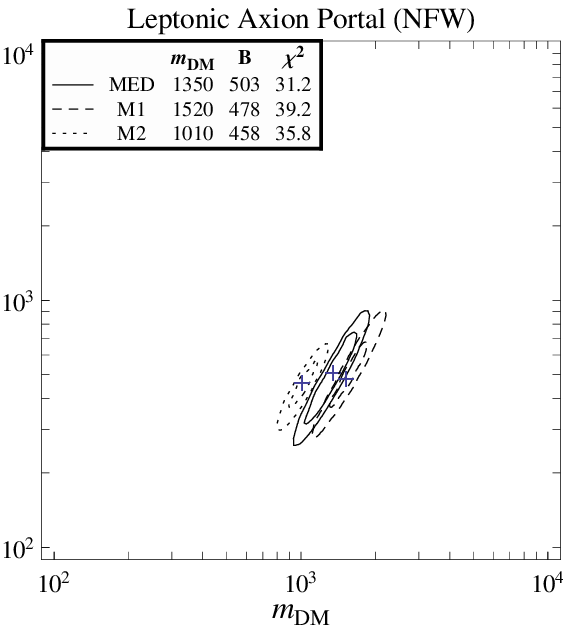}
  \includegraphics[scale=0.98]{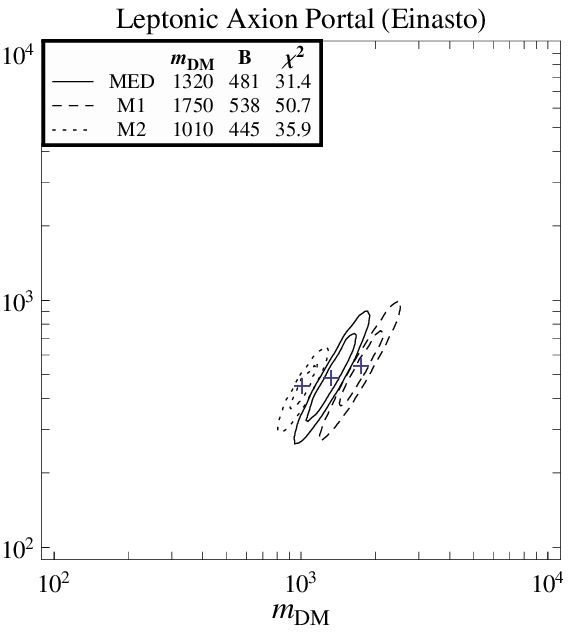}}
\caption{The best fit regions for the dark matter mass $m_{\rm DM}$ 
 and boost factor $B$ in the minimal axion portal (top row, 
 $a \rightarrow \mu^+ \mu^-$) and leptonic axion portal (bottom 
 row, $a_\ell \rightarrow e^+ e^-$) for different halo profiles 
 and propagation models.  The best fit values are indicated by the 
 crosses, and the contours are for $1\sigma$ and $2 \sigma$.}
\label{fig:positron_axion}
\end{figure}
\begin{figure}
  \center{\includegraphics[scale=0.98]{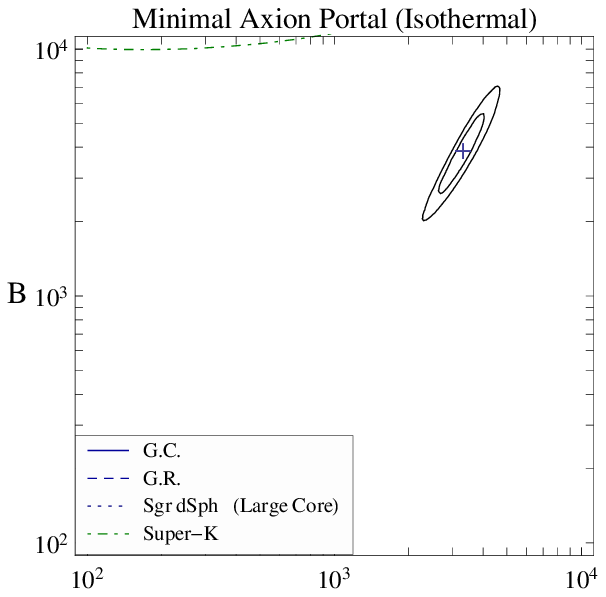}
  \includegraphics[scale=0.98]{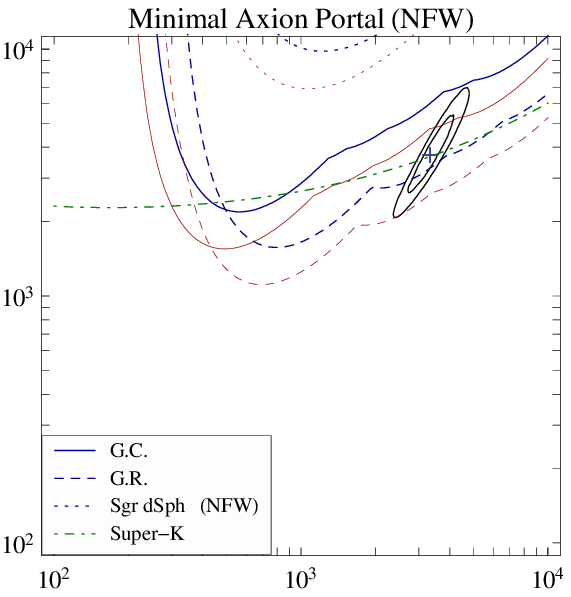}
  \includegraphics[scale=0.98]{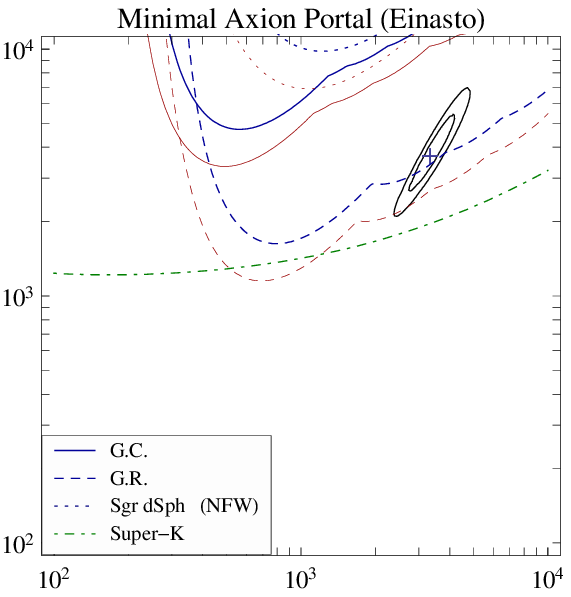}}
  \center{\includegraphics[scale=0.98]{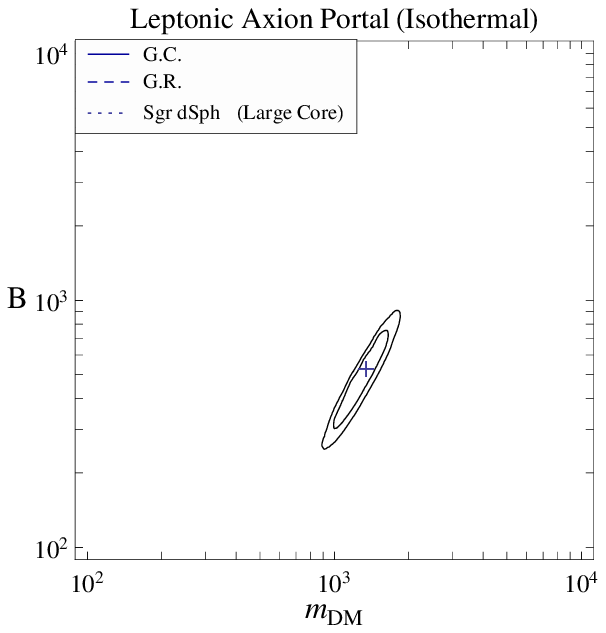}
  \includegraphics[scale=0.98]{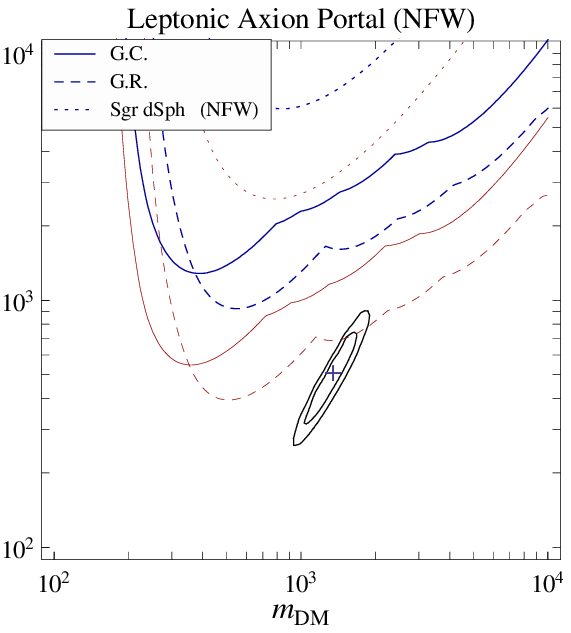}
  \includegraphics[scale=0.98]{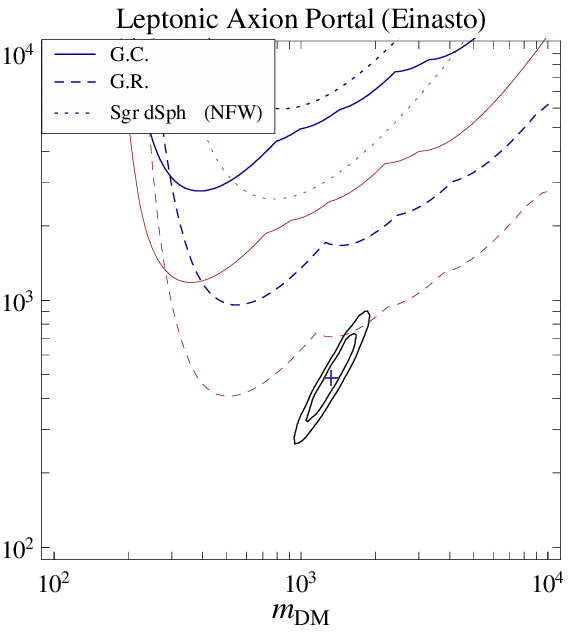}}
\caption{Constraints from gamma ray, GC (solid), GR (dashed), 
 and Sgr~dSph (dotted), and neutrino (dot-dashed) observations 
 in the $m_{\rm DM}$-$B$ plane in the minimal axion portal (top row, 
 $a \rightarrow \mu^+ \mu^-$) and leptonic axion portal (bottom row, 
 $a_\ell \rightarrow e^+ e^-$).  All the constraints, as well as the 
 best fit region for PAMELA/ATIC (MED propagation), are plotted assuming 
 that $B$ is common.  Each of the GC, GR, and Sgr~dSph constraints 
 consist of two curves.  For the minimal axion portal, the upper 
 (blue) curve is $m_a = 600~{\rm MeV}$ and the lower (red) curve 
 is $m_a = 1~{\rm GeV}$.  For the leptonic axion portal, the upper 
 (blue) curve is $m_{a_\ell} = 10~{\rm MeV}$ and the lower (red) 
 curve is $m_{a_\ell} = 100~{\rm MeV}$, which differs from the 
 choice in Figure~\ref{fig:constraint_e}.  Note that the constraint 
 lines in the cored isothermal case are above the plot region, and 
 that the halo profiles for Sgr~dSph are given in the legends.}
\label{fig:constraint_axion}
\end{figure}
\begin{figure}
  \center{\includegraphics[scale=0.65]{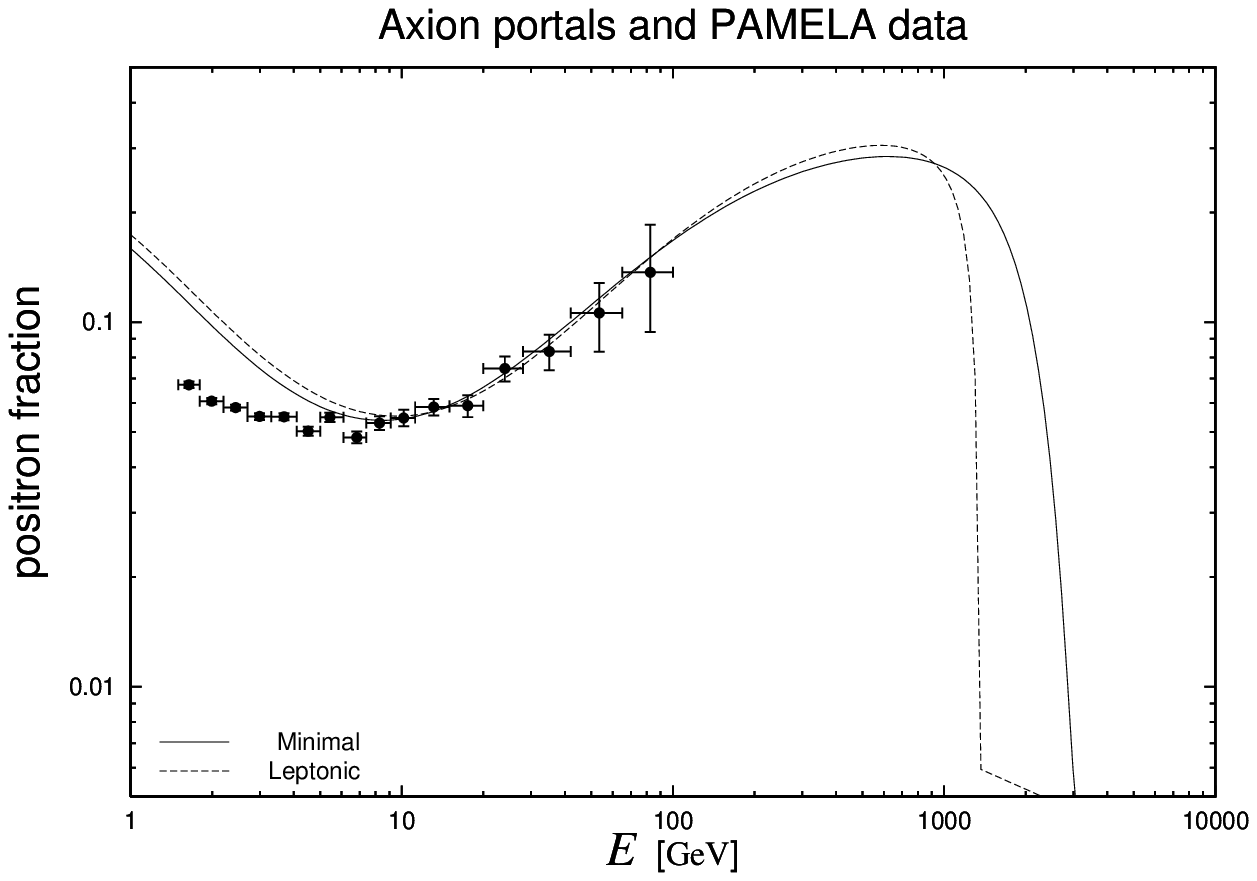}
  \hspace{0.5cm}
  \includegraphics[scale=0.65]{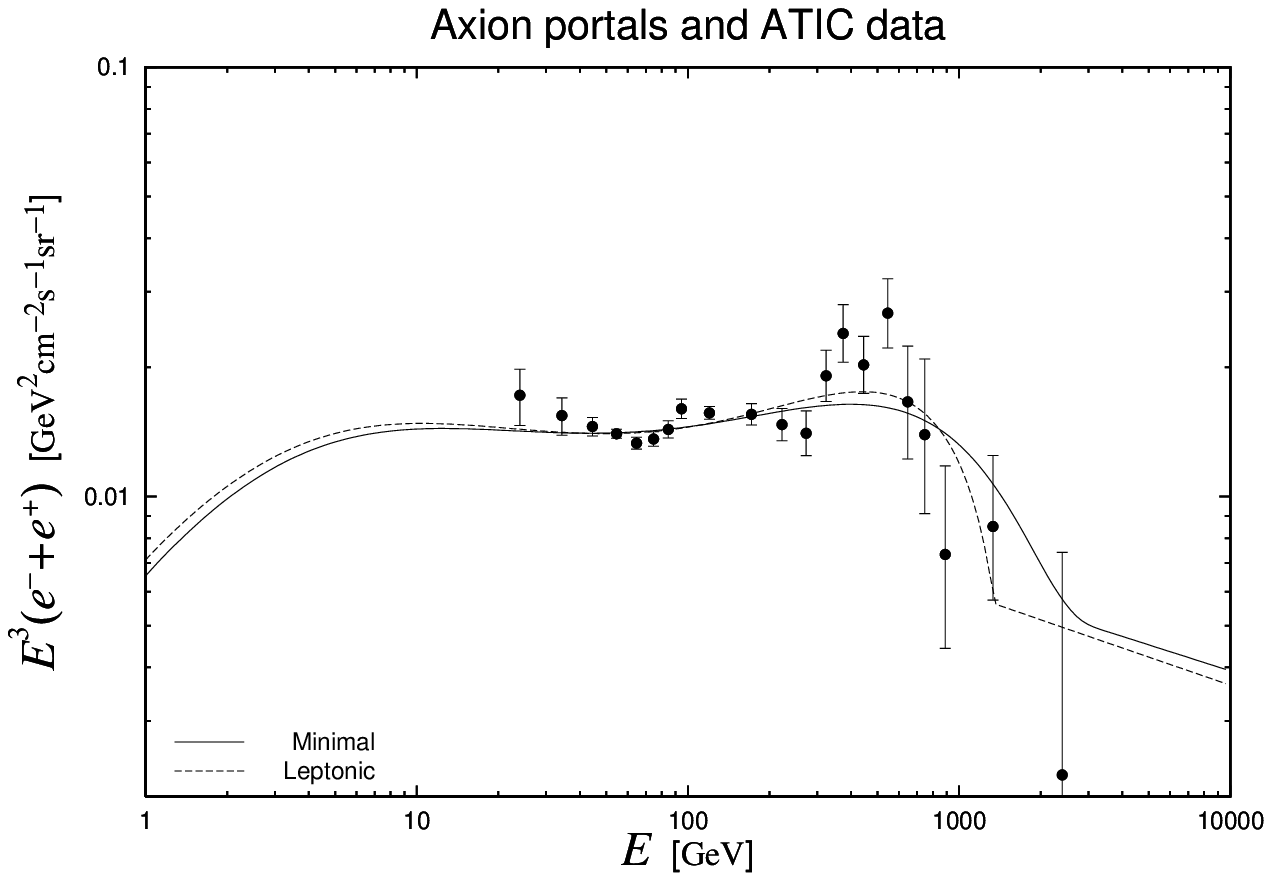}}
\caption{The predicted $e^\pm$ intensities compared to the PAMELA (left) 
 and ATIC (right) data for the minimal axion portal (solid, $a \rightarrow 
 \mu^+ \mu^-$) and leptonic axion portal (dashed, $a_\ell \rightarrow 
 e^+ e^-$).  The NFW halo profile and the MED propagation model are 
 chosen, and the $e^\pm$ backgrounds are marginalized as described 
 in Eq.~(\ref{eq:marginalize}).  Note that we fit the PAMELA data 
 only for $E \simgt 10~{\rm GeV}$ because solar modulation effects 
 are important at lower energies.}
\label{fig:spectra_axion}
\end{figure}

In Figure~\ref{fig:positron_axion}, we show the best fit values 
for $m_{\rm DM}$ and $B$ for the PAMELA/ATIC data for the three 
different diffusion models and three different halo profiles.  In 
Figure~\ref{fig:constraint_axion}, we compare the best fit regions to 
FSR and neutrino constraints.  Here we assume that $B$ is common for 
electron, gamma ray, and neutrino signals, so that the same qualifications 
as Figures~\ref{fig:constraint_e} and \ref{fig:constraint_mu} apply. 
To account for the fact that smaller masses for $a_\ell$ are 
allowed, we are considering smaller values of $m_{a_\ell}$ 
in Figure~\ref{fig:constraint_axion} than those of $m_1$ in 
Figure~\ref{fig:constraint_e}.  For completeness, we also 
show the best fit spectra to the PAMELA and ATIC data in 
Figure~\ref{fig:spectra_axion}, and the $p$-values of the 
fit in Table~\ref{tab:p-axion}.  As expected, the best fit 
values and qualitative features of the plots are partway 
between a $1$-step and a $2$-step cascade scenario.
\begin{table}
\begin{center}
\begin{tabular}{c|c|c}
  & Minimal Axion Portal & Leptonic Axion Portal 
\\ \hline
  Isothermal & $7.9\%$ & $9.2\%$ \\
  NFW        & $7.9\%$ & $9.2\%$ \\
  Einasto    & $8.3\%$ & $8.8\%$ 
\end{tabular}
\end{center}
\caption{The $p$-values for the best propagation model for each plot 
 in Figure~\ref{fig:positron_axion}.}
\label{tab:p-axion}
\end{table}

Depending on the axion mass, there are also potential gamma ray constraints 
from rare $a \rightarrow \gamma \gamma$ and $a \rightarrow \pi^+ \pi^- \pi^0$ 
decay modes, and we can place bounds on the branching fractions to these 
modes.  The $a \rightarrow \gamma \gamma$ mode is potentially dangerous 
for the entire axion mass range, but the $a \rightarrow \pi^+ \pi^- \pi^0$ 
mode only when $m_a > 3 m_\pi$.  The gamma ray spectra for these rare decays 
are given in Appendix~\ref{subapp:axion}.  For each halo profile and for 
the three H.E.S.S. data sets, we turn off FSR and find the values of the 
branching ratios where the H.E.S.S. bound is saturated for the best fit 
values of $m_{\rm DM}$ and $B$ and the best propagation model (assuming 
$B$ is common for $e^\pm$ and gamma rays).

The branching ratio bounds are shown in Tables~\ref{tab:min-axion-Br} 
and \ref{tab:lepto-axion-Br}.  The constraints from Sgr~dSph match 
the estimates in~\cite{Nomura:2008ru}, but the GC and GR data imply 
a factor of $5$--$10$ stronger bound.  In the minimal axion portal, 
${\rm Br}(a \rightarrow \gamma \gamma) \sim 10^{-5}$--$10^{-4}$, 
which is safe from the bounds.  There is somewhat more tension for 
$a \rightarrow \pi^+ \pi^- \pi^0$ where the estimated branching fraction 
was $\sim 10^{-2}$.  Note, however, that the constraints from the GC 
and GR data are weak for any halo profile relatively flat within about 
$100~{\rm pc}$ of the galactic center.  In particular, there is no 
meaningful branching ratio constraint for the cored isothermal profile 
in Eq.~(\ref{eq:rho-Isothermal}).
\begin{table}
\begin{center}
\begin{tabular}{c|ccc}
  ${\rm Br}(a \rightarrow \gamma \gamma)$ & GC & GR & Sgr~dSph 
\\ \hline
  NFW Best     & $2.2 \cdot 10^{-3}$ & $1.1 \cdot 10^{-3}$ 
    & $2.5 \cdot 10^{-2}$ \\
  Einasto Best & $4.8 \cdot 10^{-3}$ & $1.2 \cdot 10^{-3}$ 
    & --- 
\end{tabular}

\vspace{0.5cm}

\begin{tabular}{c|ccc}
  ${\rm Br}(a \rightarrow \pi^+ \pi^- \pi^0)$ & GC & GR & Sgr~dSph 
\\ \hline
  NFW Best     & $1.3 \cdot 10^{-2}$ & $7.5 \cdot 10^{-3}$ 
    & $5.2 \cdot 10^{-2}$ \\
  Einasto Best & $3.0 \cdot 10^{-2}$ & $7.9 \cdot 10^{-3}$ 
    & --- 
\end{tabular}
\end{center}
\caption{Bounds from gamma rays on the branching fractions of 
 $a \rightarrow \gamma \gamma$ and $a \rightarrow \pi^+ \pi^- \pi^0$ 
 in the minimal axion portal ($a \rightarrow \mu^+ \mu^-$).  These 
 are obtained neglecting all other sources of gamma rays and correspond 
 to the best fit values for $m_{\rm DM}$ and $B$ and the propagation 
 model giving smallest $\chi^2$.  The bounds assume an equal boost 
 factor for $e^\pm$ and gamma rays, and should be multiplied by 
 $B_{e,\mathrm{astro}}/B_{\gamma,\mathrm{astro}}$ if the boost 
 factors differ.}
\label{tab:min-axion-Br}
\end{table}
\begin{table}
\begin{center}
\begin{tabular}{c|ccc}
  ${\rm Br}(a_\ell \rightarrow \gamma \gamma)$ & GC & GR & Sgr~dSph 
\\ \hline
  NFW Best     & $8.2 \cdot 10^{-3}$ & $4.6 \cdot 10^{-3}$ 
    & $6.5 \cdot 10^{-2}$ \\
  Einasto Best & $1.8 \cdot 10^{-2}$ & $4.9 \cdot 10^{-3}$ 
    & --- 
\end{tabular}
\end{center}
\caption{The same as Table~\ref{tab:min-axion-Br}, but for the leptonic 
 axion portal ($a_\ell \rightarrow e^+ e^-$).  The $a \rightarrow 
 \pi^+ \pi^- \pi^0$ mode is irrelevant in this case.}
\label{tab:lepto-axion-Br}
\end{table}

Gamma ray and radio bounds on the axion portal were also considered 
in Ref.~\cite{Bergstrom:2008ag}, where potential $s \rightarrow b\bar{b}$ 
and $s \rightarrow \tau^+ \tau^-$ decay modes were included.  Since 
these additional $s$ decay modes depend on details of the model, we do 
not consider them here.  For FSR gamma rays, our results and conclusions 
agree qualitatively with~\cite{Bergstrom:2008ag}.  As mentioned already, 
we do not consider $408~{\rm MHz}$ radio observations to place any 
meaningful constraint on the axion portal.

\section{Conclusions}
\label{sec:concl}

The possibility of indirect detection of dark matter has been considered 
for over 25 years~\cite{Zeldovich:1980st}, but the annihilation rates 
expected from WIMP thermal relics are typically too small to give 
appreciable gamma ray or neutrino fluxes from the galactic center 
unless a very peaked dark matter halo profile is assumed.  If the 
PAMELA/ATIC data is indicative of dark matter annihilation, however, 
then the galactic annihilation rate must be boosted by $O(1000)$. 
This large boost factor considerably enhances the potential for galactic 
gamma ray and neutrino signals from the dark sector.  In this context, 
hints from the WMAP Haze may also point towards an annihilation 
explanation of PAMELA/ATIC.

In this paper, we have explored the robustness of dark matter annihilation 
predictions by considering cascade scenarios where dark matter annihilates 
into new resonances that in turn decay in one or more steps into standard 
model leptons.  These cascade annihilation scenarios are directly 
motivated by the PAMELA/ATIC data, since light resonances can enhance 
the galactic annihilation rate through nonperturbative effects and 
explain the lepton-richness of the annihilation through kinematic 
thresholds.

We have shown that electron and muon cascades give reasonable fits to 
the PAMELA/ATIC data.  As a rule of thumb, the best fit dark matter mass 
and boost factor both scale as $2^n$ for $n$-step cascade decays.  We 
then compared these best fit values to constraints from gamma rays and 
neutrinos.  The gamma ray bounds from FSR can be weakened by an order 
of magnitude through cascade decays, although increasing the length of 
cascades does not further weaken the bounds.  Neutrino bounds for dark 
matter annihilating into muons are robust to changing the length of 
the cascade, which is particularly relevant for models with large 
branching fractions to muons such as the minimal axion portal.

Assuming standard NFW or Einasto halo profiles, there is tension 
between a dark matter annihilation interpretation of PAMELA/ATIC and 
the non-observation of galactic gamma rays or neutrinos.  Such tension 
does not invalidate a dark matter annihilation hypothesis since there 
is considerable uncertainty in the dark matter halo distribution 
and velocity profile, and the constraints are uniformly weaker for 
shallower halo profiles.  For gamma rays in particular, the galactic 
center and galactic ridge constraints assume an understanding of the 
dark matter halo profile in the inner $100~{\rm pc}$ of the galaxy, 
where there is considerable uncertainty.  The dark matter halo profile 
in the inner $4''$ of the galaxy is even more uncertain, so we do not 
consider radio measurements of synchrotron to be constraining.  Also, 
for both gamma rays and neutrinos, the bounds can be weakened if the 
astrophysical boost factor for electrons/positrons is larger than those 
for gamma rays and neutrinos.

If a dark matter annihilation scenario is realized in nature with 
the boost factor suggested by PAMELA/ATIC, then one would expect 
future experiments to see a gamma ray or neutrino flux given 
standard halo assumptions.  ANTARES~\cite{Aslanides:1999vq}, 
IceCube~\cite{Ahrens:2003ix}, and KM3NeT~\cite{Kappes:2007ci} 
will greatly increase current sensitivity to upward-going muons 
resulting from galactic neutrinos.  Future atmospheric Cerenkov 
telescopes as envisioned in Ref.~\cite{Buckley:2008ud} will also 
improve the prospects of finding gamma rays from dark matter 
annihilation.  While we did not include the effect of ICS in our 
gamma ray analysis, ICS is expected to be a dominant dark matter 
annihilation signal in the energy range available to the Fermi 
Gamma-ray Space Telescope~\cite{FERMI}.  Ultimately, one hopes 
that future experiments could probe the detailed energy spectra 
of dark matter annihilation products to distinguish between 
direct annihilation and the cascade scenarios considered here.

\paragraph{\bf Note added:} While completing this paper, we became 
aware of Ref.~\cite{MPV} which considers similar issues.

\section*{Acknowledgments}

We thank Shantanu Desai from the Super-K collaboration for providing 
us with $95\%$ confidence bounds on the upward-going muon flux.  This 
work was supported in part by the Director, Office of Science, Office 
of High Energy and Nuclear Physics, of the US Department of Energy under 
Contract DE-AC02-05CH11231, and in part by the National Science Foundation 
under grant PHY-0457315.  The work of Y.N. was supported by the National 
Science Foundation under grant PHY-0555661, by a DOE OJI, and by the 
Alfred P. Sloan Foundation.  D.S. is supported by the National Science 
Foundation, and J.T. is supported by the Miller Institute for Basic 
Research in Science.

\appendix

\section{Cascade Energy Spectra}
\label{app:spectra}

In this appendix, we present formulae for the energy spectra used in the 
text.  In general, the energy spectra of final state particles in cascade 
annihilations are functions of all the intermediate masses and helicities. 
In the limit of large mass hierarchies and scalar decays, however, the 
energy spectra greatly simplify, and we use these simplified formulae 
in our analysis.

Consider cascading fields $\phi_i$ of mass $m_i$ ($m_{i+1} > 2m_i$) and 
a final state $\psi$ with mass $m_\psi$.  Cascade annihilation occurs 
through $\phi_{i+1} \rightarrow \phi_i \phi_i$ ($i=1,2,\cdots$), and 
in the last stage, $\phi_1$ decays into $\psi + X$.  Let the energy of 
$\psi$ in the $\phi_1$ rest frame be $E_0$.  Defining
\begin{equation}
  x_0 = \frac{2 E_0}{m_1},
\qquad
  \epsilon_0 = \frac{2 m_\psi}{m_1},
\label{eq:app-def-x0-e0}
\end{equation}
the $\psi$ energy spectrum is a function of $x_0$ and $\epsilon_0$
\begin{equation}
  \frac{d \tilde{N}_\psi}{d x_0} 
  = \frac{d \tilde{N}_\psi}{d x_0}(x_0,\epsilon_0),
\label{eq:app-dN-dx0}
\end{equation}
where $\epsilon_0 \leq x_0 \leq 1$.  In the case where dark matter 
$\chi$ annihilates directly into $\psi + X$, we can regard $\phi_1$ 
as the initial state of dark matter annihilation, $\chi\chi$.  In 
this case $d \tilde{N}_\psi/d x_0$ is the primary injection spectrum 
with $m_1 = 2 m_{\rm DM}$.

Now consider the previous step in the cascade annihilation, $\phi_2 
\rightarrow \phi_1 \phi_1$, with {\it one of the $\phi_1$} decaying into 
$\psi + X$.  Let the energy of $\psi$ in the $\phi_2$ rest frame be $E_1$ 
and define
\begin{equation}
  x_1 = \frac{2 E_1}{m_2},
\qquad
  \epsilon_1 = \frac{2 m_1}{m_2}.
\label{eq:app-def-x1-e1}
\end{equation}
Assuming isotropic scalar decays, the $\psi$ energy spectrum in the 
$\phi_2$ rest frame is
\begin{equation}
  \frac{d \tilde{N}_\psi}{d x_1} = \int_{-1}^1\! d\cos\theta 
    \int_{\epsilon_0}^1\! d x_0\, \frac{d \tilde{N}_\psi}{d x_0}\, 
    \delta\left(2x_1 - x_0 - \cos\theta \sqrt{x_0^2 - \epsilon_0^2} 
    \sqrt{1 - \epsilon_1^2} \right),
\label{eq:app-master-conv}
\end{equation}
where $\theta$ is the angle between the $\psi$ momentum and the $\phi_1$ 
boost axis as measured in the $\phi_1$ rest frame.

Equation~(\ref{eq:app-master-conv}) is complicated to solve in general, 
but in the limit $\epsilon_i \rightarrow 0$ ($i=0,1,\cdots$), it reduces 
to a simple convolution:
\begin{equation}
  \frac{d \tilde{N}_\psi}{d x_1} = \int_{x_1}^1\! 
    \frac{d x_0}{x_0}\, \frac{d \tilde{N}_\psi}{d x_0} 
    + {\cal O}(\epsilon_i^2),
\label{eq:app-simple-conv-prim}
\end{equation}
where $0 \leq x_1 \leq 1$ up to ${\cal O}(\epsilon_i^2)$ effects.  This 
convolution can be iterated as many times as necessary to build up the 
desired energy spectrum for an $n$-step cascade decay:
\begin{equation}
  \frac{d \tilde{N}_\psi}{d x_n}  = \int_{x_n}^1\! 
    \frac{d x_{n-1}}{x_{n-1}}\, \frac{d \tilde{N}_\psi}{d x_{n-1}} 
    + {\cal O}(\epsilon_i^2),
\label{eq:app-simple-conv}
\end{equation}
where $x_{n-1} = 2 E_{n-1}/m_n$ with $E_{n-1}$ being the energy of 
$\psi$ in the $\phi_n$ rest frame, and $0 \leq x_n \leq 1$ up to 
${\cal O}(\epsilon_i^2)$ effects.  Note that we here adopt the 
normalization convention of
\begin{equation}
  \int_0^1\! d x_n\, \frac{d \tilde{N}_\psi}{d x_n} = 1,
\label{eq:app-norm-N}
\end{equation}
regardless of the value of $n$, so that the injection spectra per dark 
matter annihilation must be multiplied by the multiplicity of $\psi$ 
in the final state.

\subsection{Direct electron spectra}
\label{subapp:direct-e}

Here we derive the spectra of electrons/positrons arising directly 
from $\phi_1$ decay, $\phi_1 \rightarrow e^+ e^-$ (or dark matter 
annihilation, $\chi\chi \rightarrow e^+ e^-$).  Ignoring the effect 
of final state radiation to smooth the spectrum, the electron energy 
spectrum is given by
\begin{equation}
  \frac{d \tilde{N}_e}{d x_0} = \delta(1-x_0),
\label{eq:app-dNe_dx0}
\end{equation}
where we have adopted the convention $\int_0^1\! d x_0\, \delta(1-x_0) = 1$. 
The positron energy spectrum is identical.

Applying the simplified convolution formula in Eq.~(\ref{eq:app-simple-conv}) 
for an $n$-step cascade annihilation, we then find
\begin{eqnarray}
  \frac{d \tilde{N}_e}{d x_1} &=& 1,
\label{eq:app-dNe_dx1}\\
  \frac{d \tilde{N}_e}{d x_2} &=& \ln\frac{1}{x_2},
\label{eq:app-dNe_dx2}\\
  \frac{d \tilde{N}_e}{d x_n} &=& Q_n(x_n),
\label{eq:app-dNe_dxn}
\end{eqnarray}
where we have defined
\begin{equation}
  Q_n(x) \equiv \frac{1}{(n-1)!} \left( \ln\frac{1}{x} \right)^{n-1}.
\label{eq:app-Qn}
\end{equation}
Note that these are energy spectra for {\it one} of the electrons (or 
positrons), so that the electron (or positron) injection spectra per 
dark matter annihilation $\chi\chi \rightarrow 2\phi_n \rightarrow 
\cdots \rightarrow 2^n (e^+ e^-)$ are
\begin{equation}
  \frac{d N_e}{d x_n} = 2^n\, \frac{d \tilde{N}_e}{d x_n},
\label{eq:app-e-spectra}
\end{equation}
where $x_n = E_e/m_{\rm DM}$ with $E_e$ being the electron (positron) 
energy in the center-of-mass frame for dark matter annihilation.  The 
direct annihilation case, $\chi\chi \rightarrow e^+ e^-$, corresponds 
to $n=0$.

\subsection{Electron and neutrino spectra from muon decay}
\label{subapp:muon-decay}

Here we discuss the spectra of electrons, positrons and neutrinos 
arising from muon decay.  Consider $\phi_1 \rightarrow \mu^+ \mu^-$ 
(or $\chi\chi \rightarrow \mu^+ \mu^-$) followed by $\mu \rightarrow 
e \nu_e \nu_\mu$.  (One of the neutrinos here should be an anti-neutrino. 
We omit the particle-antiparticle identification here and below.) 
Assuming a massless electron, the (unpolarized) spectra of electrons 
and neutrinos in the rest frame of the muon are
\begin{eqnarray}
  && \frac{d \tilde{N}_{\mu \rightarrow e}}{d x_{-1}} 
  \,\,=\,\, \frac{d \tilde{N}_{\mu \rightarrow \nu_\mu}}{d x_{-1}} 
  \,\,=\,\, 6 (x_{-1})^2 - 4 (x_{-1})^3,
\label{eq:app-mu-e-nu_mu}\\
  && \frac{d \tilde{N}_{\mu \rightarrow \nu_e}}{d x_{-1}} 
  \,\,=\,\, 12 (x_{-1})^2 - 12 (x_{-1})^3,
\label{eq:app-mu-nu_e}
\end{eqnarray}
where we are using the notation $x_{-1} = 2E_{-1}/m_\mu$, with $E_{-1}$ 
being the energy in the muon rest frame.

Applying the cascade convolution for electrons and muon neutrinos
\begin{eqnarray}
  && \frac{d \tilde{N}_{\mu \rightarrow e}}{d x_{0}} 
  \,\,=\,\, \frac{d \tilde{N}_{\mu \rightarrow \nu_\mu}}{d x_{0}} 
  \,\,=\,\, \frac{5}{3} - 3 x_0^2 + \frac{4}{3}x_0^3,
\label{eq:app-mu-e-0}\\
  && \frac{d \tilde{N}_{\mu \rightarrow e}}{d x_{1}} 
  \,\,=\,\, \frac{d \tilde{N}_{\mu \rightarrow \nu_\mu}}{d x_{1}} 
  \,\,=\,\, -\frac{19}{18} + \frac{3}{2} x_1^2 - \frac{4}{9} x_1^3 
    + \frac{5}{3} Q_2(x_1),
\label{eq:app-mu-e-1}\\
  && \frac{d \tilde{N}_{\mu \rightarrow e}}{d x_{2}} 
  \,\,=\,\, \frac{d \tilde{N}_{\mu \rightarrow \nu_\mu}}{d x_{2}} 
  \,\,=\,\, \frac{65}{108} - \frac{3}{4} x_2^2 + \frac{4}{27} x_2^3 
    - \frac{19}{18} Q_2(x_2) + \frac{5}{3} Q_3(x_2),
\label{eq:app-mu-e-2}
\end{eqnarray}
and for electron neutrinos
\begin{eqnarray}
  && \frac{d \tilde{N}_{\mu \rightarrow \nu_e}}{d x_{0}} 
  \,\,=\,\, 2 - 6 x_0^2 + 4 x_0^3,
\label{eq:app-mu-nu_e-0}\\
  && \frac{d \tilde{N}_{\mu \rightarrow \nu_e}}{d x_{1}} 
  \,\,=\,\, -\frac{5}{3} + 3 x_1^2 - \frac{4}{3}x_1^3 
    + 2 Q_2(x_1),
\label{eq:app-mu-nu_e-1}\\
  && \frac{d \tilde{N}_{\mu \rightarrow \nu_e}}{d x_{2}} 
  \,\,=\,\, \frac{19}{18} - \frac{3}{2}x_2^2 + \frac{4}{9}x_2^3 
    - \frac{5}{3} Q_2(x_2) + 2 Q_3(x_2).
\label{eq:app-mu-nu_e-2}
\end{eqnarray}
Again, these are energy spectra for {\it one} of the electrons, positrons 
or (anti-)neutrinos.  To obtain the injection spectra per dark matter 
annihilation, we must multiply the multiplicity factor, $2^n$ for $n$-step, 
and set $x_n = E/m_{\rm DM}$.  Here, $E$ is the energy of a particle in 
the center-of-mass frame for dark matter annihilation.

For comparison, the corresponding formulae in the approximation of 
isotropic three-body decays are
\begin{eqnarray}
  && \frac{d \tilde{N}_{\mu \rightarrow e}}{d x_{-1}} 
  \,\,\simeq\,\, 2x_{-1},
\label{eq:app-mu-iso-1}\\
  && \frac{d \tilde{N}_{\mu \rightarrow e}}{d x_n} 
  \,\,\simeq\,\, (-1)^{n+1} \left( 2 x_n + 2 \sum_{i=1}^{n+1} 
    (-1)^i Q_i(x_n) \right)
  \,\,\equiv\,\, \overline{Q}_{n+1}(x_n).
\label{eq:app-mu-iso-2}
\end{eqnarray}
As a rough rule of thumb, the electron spectrum for an $n$-step muon 
cascade has a shape between  $(n+1)$- and $(n+2)$-step electron cascades.

\subsection{Gamma ray spectra from final state radiation}
\label{subapp:FSR}

Primary gamma rays come from final state radiation in the decay 
$\phi_1 \rightarrow \ell^+ \ell^- \gamma$.  In principle, one could 
do an exact calculation to ${\cal O}(\alpha_{\rm EM})$ of the gamma 
ray spectrum, which would have the full $\epsilon_0 \equiv 2m_\ell/m_1$ 
dependence.  Since we are using the simplified convolution formula 
in Eq.~(\ref{eq:app-simple-conv}), it is not consistent to keep 
${\cal O}(\epsilon_0^2)$ corrections in the exact gamma ray calculation, 
and it suffices to use twice the Altarelli-Parisi splitting formula
\begin{equation}
  \frac{d \tilde{N}_\gamma}{d x_0} 
  = \frac{\alpha_{\rm EM}}{\pi} \frac{1 + (1-x_0)^2}{x_0} 
    \left\{ -1 + \ln\left(\frac{4(1-x_0)}{\epsilon_0^2}\right) \right\},
\label{eq:app-dNgamma-dx0}
\end{equation}
where the normalization of $\tilde{N}_\gamma$ is such that $\int\! d x_n\, 
d \tilde{N}_\gamma/d x_n$ gives the (average) number of photons per 
$\phi_1$ decay.  Note that the expression of Eq.~(\ref{eq:app-dNgamma-dx0}) 
becomes negative at $x_0 > 1-e\, \epsilon_0^2/4$, which does not 
correspond to the kinematic threshold.  The error from this, however, 
is formally an ${\cal O}(\epsilon_0^2)$ effect.

Applying the simplified convolution formula, we obtain
\begin{eqnarray}
  \frac{d \tilde{N}_\gamma}{d x_1} 
  &=& \frac{\alpha_{\rm EM}}{\pi} \frac{1}{x_1} 
    \left\{ \left(-1 + \ln\frac{4}{\epsilon_0^2}\right) R_1(x_1) 
    + S_1(x_1) \right\},
\label{eq:app-dNgamma-dx1}\\
  \frac{d \tilde{N}_\gamma}{d x_2} 
  &=& \frac{\alpha_{\rm EM}}{\pi} \frac{1}{x_2} 
    \left\{ \left(-1 + \ln\frac{4}{\epsilon_0^2}\right) R_2(x_2) 
    + S_2(x_2) \right\},
\label{eq:app-dNgamma-dx2}
\end{eqnarray}
where
\begin{eqnarray}
  R_1(x) &=& 2 - x - x^2 + 2x \ln x,
\label{eq:app-R1}\\
  R_2(x) &=& 2 - 3x + x^2 + x \ln x - x (\ln x)^2,
\label{eq:app-R2}\\
  S_1(x) &=& \biggl(\frac{\pi^2}{3} - 1\biggr)x + x^2 + 2x \ln x 
    + (2 - x - x^2) \ln(1-x) - 2x\, {\rm Li}_2(x),
\label{eq:app-S1}\\
  S_2(x) &=& \biggl(\frac{\pi^2}{6} + 2 - 2\zeta(3)\biggr)x - 2x^2 
    - \biggl(\frac{\pi^2}{3} - 3\biggr) x \ln x + (2 - 3x + x^2) \ln(1-x) 
\nonumber\\
  && - x (\ln x)^2 - x\, {\rm Li}_2(x) + 2x\, {\rm Li}_3(x).
\label{eq:app-S2}
\end{eqnarray}
The photon injection spectra {\it per dark matter annihilation} are 
then given by
\begin{equation}
  \frac{d N_\gamma}{d x_n} = 2^n\, \frac{d \tilde{N}_\gamma}{d x_n},
\label{eq:app-photon-spectra}
\end{equation}
where $x_n = E_\gamma/m_{\rm DM}$, with $E_\gamma$ being the photon 
energy in the center-of-mass frame for dark matter annihilation.  Here, 
$\epsilon_0 = m_\ell/m_{\rm DM}$ for direct annihilation and $\epsilon_0 
= 2 m_\ell/m_1$ otherwise.

For the hardest gamma rays near $x_n \rightarrow 1$, the behavior of 
$d \tilde{N}_\gamma/d x_n$ is
\begin{eqnarray}
  \frac{d \tilde{N}_\gamma}{d x_0} 
  &\simeq& \frac{\alpha_{\rm EM}}{\pi} \frac{1}{x_0} 
    \ln \left( \frac{4(1-x_0)}{\epsilon_0^2} \right),
\label{eq:app-dNgamma-dx0-lim}\\
  \frac{d \tilde{N}_\gamma}{d x_1} 
  &\simeq& \frac{\alpha_{\rm EM}}{\pi} \frac{1-x_1}{x_1} 
    \ln \left(\frac{4(1-x_1)}{\epsilon_0^2} \right),
\label{eq:app-dNgamma-dx1-lim}\\
  \frac{d \tilde{N}_\gamma}{d x_2} 
  &\simeq& \frac{\alpha_{\rm EM}}{\pi} \frac{(1 - x_2)^2}{2 x_2} 
    \ln \left(\frac{4(1-x_2)}{\epsilon_0^2}\right).
\label{eq:app-dNgamma-dx2-lim}
\end{eqnarray}
Compared to direct annihilation into leptons, a $1$-step cascade 
annihilation gives a gamma ray spectrum that is suppressed not only 
by $\ln(2 m_\ell/m_1)/\ln(m_\ell/m_{\rm DM})$ but also by an additional 
suppression factor of $(1-x)$ for the highest energy gamma rays.

\subsection{Gamma ray subtlety for muons}
\label{subapp:gamma-mu}

There are actually two contributions to the gamma ray spectrum for 
$\phi_1 \rightarrow \mu^+ \mu^- \gamma$.  In addition to final state 
radiation from muons, there is the radiative decay of the muon 
$\mu \rightarrow e \nu_e \nu_\mu \gamma$.  Formally, this contribution 
is suppressed by a factor of $1/\ln(m_\mu/m_1)$ or $(1-x_n)^2$, but 
for $m_1 \approx m_\mu$, it is an important effect.

The gamma ray spectrum in the muon rest frame is known in the limit 
that $r= m_e^2/m_\mu^2$ is small~\cite{Kuno:1999jp}.   Assuming 
unpolarized muons, we can derive the $0$-step cascade annihilation 
spectrum from the muon rest frame spectrum
\begin{eqnarray}
  \frac{d \tilde{N}_{\mu \rightarrow \gamma}}{d x_{-1}} 
  &=& \frac{\alpha_{\rm EM}}{3\pi} \frac{1}{x_{-1}} 
    \left( T_{-1}(x_{-1}) \ln \frac{1}{r} + U_{-1}(x_{-1}) \right),
\label{eq:app-dNmugamma_dx-1}\\
  \frac{d \tilde{N}_{\mu \rightarrow \gamma}}{d x_{0}} 
  &=& \frac{\alpha_{\rm EM}}{3\pi} \frac{1}{x_{0}} 
    \left( T_{0}(x_{0}) \ln \frac{1}{r} + U_{0}(x_{0}) \right),
\label{eq:app-dNmugamma_dx0}
\end{eqnarray}
where
\begin{eqnarray}
  T_{-1}(x) &=& (1-x)(3 -2x + 4x^2 - 2x^3),
\label{eq:app-T-1}\\
  T_{0}(x) &=& 3 + \frac{2}{3}x - 6 x^2 + 3x^3 
    - \frac{2}{3}x^4 + 5 x \ln x,
\label{eq:app-T0}\\
  U_{-1}(x) &=& (1-x) \biggl( -\frac{17}{2} + \frac{23}{6}x 
    - \frac{101}{12}x^2 + \frac{55}{12}x^3 
    + (3 -2x + 4x^2 - 2x^3)\ln(1-x) \biggr),
\label{eq:app-U-1}\\
  U_{0}(x) &=& -\frac{17}{2} - \frac{3}{2}x + \frac{191}{12}x^2 
    - \frac{23}{3}x^3 + \frac{7}{4}x^4 
    + \biggl( 3+\frac{2}{3}x - 6 x^2 + 3 x^3 - \frac{2}{3}x^4 \biggr) \ln (1-x)
\nonumber\\
  && - \frac{28}{3}x \ln x + 5x \ln (1-x) \ln x + 5x\, {\rm Li}_2(1-x).
\label{eq:app-U0}
\end{eqnarray}
The convolutions for $1$- and $2$-step decays are straightforward 
to derive.

Note again that $\int\! d x_n\, d \tilde{N}_{\mu \rightarrow \gamma}/d x_n$ 
give the (average) number of photons per muon decay.  The photon 
spectra from radiative muon decay {\it per dark matter annihilation} 
are then given by $2^n d \tilde{N}_{\mu \rightarrow \gamma}/d x_n$. 
The total photon injection spectra per dark matter annihilation are 
given by
\begin{equation}
  \frac{d N_\gamma}{d x_n} = 2^n \left( \frac{d \tilde{N}_\gamma}{d x_n} 
    + 2 \frac{d \tilde{N}_{\mu \rightarrow \gamma}}{d x_n} \right),
\label{eq:app-photon-mu-spectra}
\end{equation}
where $x_n = E_\gamma/m_{\rm DM}$, with $E_\gamma$ being the photon 
energy in the center-of-mass frame for dark matter annihilation.

\subsection{Rare modes in the axion portal}
\label{subapp:axion}

In Section~\ref{sec:axion}, we consider bounds on rare 
$a \rightarrow \gamma \gamma$ and $a \rightarrow \pi^+ \pi^- \pi^0$ 
decay modes in axion portal models.  The axion portal has both a $1$-step 
and a $2$-step component, and the gamma ray spectrum for each can be 
calculated straightforwardly.  The $a \rightarrow \gamma \gamma$ spectra 
are identical (up to normalization) to the $\phi_1 \rightarrow e^+ e^-$ 
spectra already calculated:
\begin{eqnarray}
  \frac{d \tilde{N}_{a \rightarrow \gamma}}{d x_1} &=& 2,
\label{eq:app-a-gamma-1}\\
  \frac{d \tilde{N}_{a \rightarrow \gamma}}{d x_2} &=& 2\ln\frac{1}{x_2},
\label{eq:app-a-gamma-2}
\end{eqnarray}
where the normalization of $\tilde{N}_{a \rightarrow \gamma}$ is such 
that $\int\! d x_n\, d \tilde{N}_{a \rightarrow \gamma}/d x_n$ gives 
the number of photons per $a$ decay.

For $a \rightarrow \pi^+ \pi^- \pi^0$ followed by $\pi^0 \rightarrow 
\gamma \gamma$, we can use the $\overline{Q}_n$ function from 
Eq.~(\ref{eq:app-mu-iso-2}) if we assume that $m_a \gg m_\pi$ 
and that the $a \rightarrow 3 \pi$ decay is isotropic:
\begin{eqnarray}
  \frac{d \tilde{N}_{a \rightarrow \pi^0 \rightarrow \gamma}}{d x_1} 
  &\simeq& 2 \overline{Q}_2(x_1),
\label{eq:app-a-pi0-1}\\
    \frac{d \tilde{N}_{a \rightarrow \pi^0 \rightarrow \gamma}}{d x_2} 
  &\simeq& 2 \overline{Q}_3(x_2),
\label{eq:app-a-pi0-2}
\end{eqnarray}
where again the normalization of $\tilde{N}_{a \rightarrow 
\pi^0 \rightarrow \gamma}$ is such that $\int\! d x_n\, 
d \tilde{N}_{a \rightarrow \pi^0 \rightarrow \gamma}/d x_n$ gives 
the number of photons per $a$ decay.  One should keep in mind that 
$m_a \simeq 3 m_\pi$ in the region of interest, but the hierarchical 
cascade approximation is still reasonably representative of the 
true energy spectrum.

\section{Leptonic Axion Portal}
\label{app:lepto-axion}

In the minimal axion portal construction, the axion has large hadronic 
couplings, and is therefore strongly constrained by beam dump and 
rare meson decay experiments.  In particular, the axion is forced to 
decay primarily into muons, and, as we saw in Section~\ref{sec:neutrino}, 
there is some degree of tension between a muon annihilation 
scenario and the absence of galactic neutrinos.  Also, we saw 
in Section~\ref{sec:axion} that there are strong gamma ray bounds 
on the $a \rightarrow \pi^+ \pi^- \pi^0$ decay mode, which in 
the minimal axion portal arises from axion-pion mixing.

Since the decay properties of the axion are irrelevant for dark matter 
freezeout, we can easily modify the couplings of the axion to standard 
model fields without losing the good features of this scenario.  In 
particular, we can construct a leptonic axion portal model where the 
axion has no hadronic couplings.  While such a leptonic axion could 
decay into muons as in the minimal axion portal, in the text we consider 
the less constrained case where the leptonic axion decays primarily 
into electrons.

The simplest example for the leptonic axion portal can be constructed 
as follows.  Vector-like fermion dark matter $\psi/\psi^c$ obtains 
a mass from spontaneous symmetry breaking through the vacuum expectation 
value of a complex scalar $S_\ell$:
\begin{equation}
  {\cal L} = - \xi S_\ell \psi \psi^c + {\rm h.c.},
\qquad
  S_\ell = \left( f_a + \frac{s_\ell}{\sqrt{2}} \right) 
    e^{i a_\ell/\sqrt{2} f_a},
\label{eq:app-lepto-axion}
\end{equation}
where $a_\ell$ is the pseudoscalar axion, $s_\ell$ is a light scalar, 
and $f_a$ is the axion decay constant, which is assumed to be of order 
TeV.  As in the minimal axion portal, the masses of $s_\ell$ and $a_\ell$ 
can be considered as free parameters.  In order for $a_\ell$ to decay 
into leptons, $S_\ell$ must be charged under a leptonic symmetry (which 
is softly broken to give a mass to $a_\ell$).  This requires introducing 
separate electron-type and neutrino-type Higgses:
\begin{equation}
  {\cal L} = -\lambda_e \ell h_e e^c - \lambda_\nu \ell h_\nu \nu^c 
    - A_\ell S_\ell h_e h_\nu + {\rm h.c.}
\label{eq:app-lepto-Yukawa}
\end{equation}
These interactions force the standard model leptons to carry axial 
leptonic charges.  (Small neutrino masses can be obtained through the 
standard see-saw mechanism, and it is straightforward to extend the 
model to incorporate supersymmetry.)

In order to eliminate the hadronic couplings of $a_\ell$, the standard 
model quarks are assumed not to carry charges under the leptonic symmetry. 
This requires introducing one or more new Higgses for the quark sector, 
which must also be singlets under the leptonic symmetry.  As long as 
the potential for these Higgses preserves the leptonic symmetry, then 
$a_\ell$ will have no hadronic couplings and cannot mix with the neutral 
mesons after the leptonic symmetry is spontaneously broken.

The absence of hadronic couplings allows $m_{a_\ell}$ to be lighter 
than $2 m_\mu$, and thus $a_\ell$ to decay dominantly into electrons 
while satisfying beam dump and rare meson decay constraints.  Also, the 
leptonic axion $a_\ell$ has no $a_\ell \rightarrow \pi^+ \pi^- \pi^0$ 
decay mode, reducing the gamma ray constraints in the case that 
$m_{a_\ell} > 3 m_\pi$ (where $a_\ell$ dominantly decays into muons). 
Since the strongest astrophysical bounds on light degrees of freedom 
come from hadronic couplings~\cite{Raffelt:1990yz}, $m_{a_\ell}$ 
might even be as light as $2 m_e$, although a detailed study of the 
constraints on the leptonic axion is beyond the scope of this paper.

As an example of the quark sector interactions, there could be separate 
up-type and down-type Higgses.  In this case, it is natural to 
assume a hadronic symmetry under which the quarks carry axial charges:
\begin{equation}
  {\cal L} = -\lambda_u q h_u u^c - \lambda_d q h_d d^c 
    - A_q S_q h_u h_d + {\rm h.c.}
\label{eq:app-hadro-Yukawa}
\end{equation}
The axion contained in the field $S_q$ could then be the QCD axion 
and solve the strong CP problem.  To avoid astrophysical constraints, 
however, the vacuum expectation value of $S_q$ must be $\simgt 
10^{10}~{\rm GeV}$, much larger than $\langle S_\ell \rangle 
= O({\rm TeV})$.  Explicit breaking of the hadronic symmetry 
must also be much smaller than that of the leptonic symmetry.

\end{document}